\documentclass{aa}  

\usepackage{graphicx}
\usepackage{txfonts}
\usepackage{natbib}
\usepackage{hyperref}
\usepackage{graphicx}
\usepackage{xcolor}
\usepackage{multirow}
\usepackage{rotating}
\defcitealias{Vazza18_magne}{V18}
\defcitealias{Dominguez19}{DF19}
\newcommand{\myzenodo}{\href{https://doi.org/10.5281/zenodo.21870521}{our online repository}}
\newcommand{\muG}{\mu \rm{G}}
\newcommand{\kms}{\rm{km\,s^{-1}}}
\newcommand{\erf}{\rm{erf}}

\newcommand{\utot}{u_{\rm{tot}}}
\newcommand{\ucomp}{u_{\rm{comp}}}
\newcommand{\usol}{u_{\rm{sol}}}
\newcommand{\rs}{r_{\rm{s}}}
\newcommand{\betatherm}{\beta_{\rm{therm}}}
\newcommand{\proton}{m_{\rm{p}}}

\newcommand{\Esol}{E_{\rm{sol}}}
\newcommand{\Ecomp}{E_{\rm{comp}}}

\newcommand{\sigmamean}{\sigma_{\rm{mean}}}
\newcommand{\sound}{c_{\rm{s}}}
\newcommand{\lF}{l_{\rm{F}}}
\newcommand{\lC}{l_{\rm{C}}}
\newcommand{\EMk}{E_{\rm{M}} (k)}
\newcommand{\EM}{E_{\rm{M}}}
\newcommand{\kny}{k_{\rm{Ny}}}
\newcommand{\Nindep}{N_{\rm{indep}}}
\newcommand{\gammamean}{\gamma_{\rm{mean}}}
\newcommand{\fkol}{f_{\rm{Kolm}}}
\newcommand{\kboltz}{k_{\rm{B}}}

\begin{document}

   \title{Monitoring the power spectrum of magnetic field fluctuations across the simulated intracluster medium}

  \author{T. Bartalesi\inst{1,2}\fnmsep\thanks{\email{tommaso.bartalesi3@unibo.it}}
          \and
          F. Vazza\inst{1,3}
          \and
          S. Ettori\inst{2, 4}
          \and
          C. Nipoti\inst{1}
          \and
          P. Dom\'inguez-Fern\'andez\inst{5,6}
          }

%\fnmsep\thanks{\email{stefano.ettori@inaf.it}}

   \institute{Dipartimento di Fisica e Astronomia “Augusto Righi” – Alma Mater Studiorum – Università di Bologna, via Gobetti 93/2, I-40129 Bologna 
         \and INAF, Osservatorio di Astrofisica e Scienza dello Spazio, via Piero Gobetti 93/3, 40129 Bologna, Italy 
         \and 
        INAF - IRA, via P.~Gobetti 101, 40129 Bologna, Italy 
         \and INFN, Sezione di Bologna, viale Berti Pichat 6/2, 40127 Bologna, Italy
        \and 
        Instituto de Ciencias Nucleares, Universidad Nacional Aut\'onoma de M\'exico, A. P. 70-543 04510 D. F. Mexico
        \and
        Center for Astrophysics $\vert$ Harvard \& Smithsonian, 60 Garden Street, Cambridge, MA 02138, USA}

\abstract
  {Thanks to the current and forthcoming radio interferometers, Faraday rotation measure maps extend out to the outskirts of galaxy clusters. Interpreted through models of the magnetic power spectrum, these maps can provide stringent constraints on the properties of the magnetic field of the intracluster medium (ICM).} %leave it empty if necessary 
   {Our goal is to ascertain how the magnetic power spectrum varies with the environment and dynamics of the ICM in a post-merger galaxy cluster formed in a state-of-the-art non-radiative cosmological magneto-hydrodynamic simulation.
   }
   {
   We divide the simulated cluster into cubic subboxes with side length 950 kpc and compute the kinetic and magnetic power spectra in each. 
   Starting from a "dynamo-only" functional form proposed in previous works to describe magnetic fluctuations amplified by the small-scale dynamo, we introduce an additional Kolmogorov-like component to account for the contribution of unamplified magnetic field patches. 
   We fit this extended model to the magnetic power spectrum of each subbox individually using a Markov chain Monte Carlo method and examine scatter plots of the inferred parameters versus the local ICM properties evaluated in the corresponding subboxes.
   }
   % Results (mandatory)
    {
    The extended model reproduces the magnetic power spectrum data well in most subboxes, whereas the "dynamo-only" model fits these data well only in subboxes not too far from the center. 
    The inferred parameters exhibit substantial subbox-to-subbox variations that correlate significantly with the local ICM density and clumpiness.
    }
    {These correlations suggest that the magnetic power spectrum is closely linked to the local ICM environment,
    and may help observers in the interpretation of Faraday Rotation data.}


\maketitle

%
%-------------------------------------------------------------------

\section{Introduction}
\label{sec:intro}

The intracluster medium (ICM), the rarefied ($10^{-1}$ -- $10^{-5}$ cm$^{-3}$) hot ($\sim$ keV) plasma permeating galaxy clusters, is known to be weakly magnetized, with a magnetic field strength up to a few $\muG$ \citep[e.g.][for a review]{vanWeeren19}.
Our theoretical understanding and our ability to appropriately model the power spectrum of the ICM magnetic field significantly contribute to the accuracy in the magnetic field strength constraints.

The observed polarization angle of a synchrotron-emitting source in the background of a cluster or embedded within the cluster itself is rotated, with respect to the intrinsic polarization angle, proportional to the observing wavelength squared, and by a quantity known as Faraday Rotation Measure (RM), $\Delta \phi \propto  \lambda^2   RM $. 
The RM is proportional to the magnetic field vector component along the line of sight (LOS).
Positive or negative rotations of the polarization angle are indeed induced according to the LOS magnetic field direction \citep[e.g.][]{Enblin03}.
The typical angular resolution of currently available radio interferometers allows us to derive the two-dimensional (2D) RM maps that resolve fluctuations down to coherence lengths of a few kpc in nearby objects \citep[see][for few examples]{Murgia04,Bonafede10,Govoni17}. 
The RM maps obtained for individual polarized radio sources less extended than 100 kpc, such as background and cluster-member radio galaxies, have been obtained mainly in the cluster inner regions.
More recently, new observations using MeerKAT\footnote{\hyperlink{https://www.sarao.ac.za/science/meerkat/}{https://www.sarao.ac.za/science/meerkat/}} and LOFAR\footnote{\hyperlink{https://www.astron.nl/telescopes/lofar/}{https://www.astron.nl/telescopes/lofar/}} have allowed us to explore the trend of RM in cluster outskirts, and more specifically in the direction of radio relics, i.e.\ $\sim$ Mpc extended, relatively highly polarized ($\approx20 - 30\%$ of total radio emission) synchrotron-emitting sources located in cluster outskirts and likely produced by merger shocks \citep[e.g.][and references therein]{vanWeeren19}.
This has improved the RM coverage of the cluster outskirts and in combination with mock RM maps produced for assumed three-dimensional models of cluster magnetic fields, has allowed to infer a profile for the magnetic field strength in the ICM.
The state-of-the-art RM studies typically assume that the shape of the magnetic power spectrum remains constant throughout the cluster volume, while allowing its normalization to vary freely. 
The spectrum normalization directly determines the magnetic field strength, and it follows from assuming a $B \propto \rho^\eta$ scaling with the ICM density, $\rho$ \citep[e.g.][]{do99}. These analyses have constrained the magnetic field strength to decrease from values of up to $\sim 10\,\mu\mathrm{G}$ in cluster cores to a few $\sim 0.1\,\mu\mathrm{G}$ in the outskirts, while the exponent $\eta$ is usually determined to be in the range (0.5 -- 1).  
The smallest scales of the magnetic fluctuations are typically inferred to be $\lesssim 10$ kpc, a value almost comparable to the spatial resolution of currently available interferometers 
\citep[e.g.,][]{Murgia04,Bonafede10,Vacca10, Bonafede13,Stuardi21,DeRubeis24,Osinga25,Loi26, Bon26}. 
Also, different assumptions on the shape of the magnetic power spectrum (either a Kolmogorov spectrum or one consistent with cosmological MHD simulations), at fixed density profile and $\eta$ can lead to differences in the inferred magnetic field strength profiles up to a factor $\sim 2$ \citep[e.g.][]{Stuardi21}.  Although this discrepancy is comparable to the uncertainty associated with the limited spatial sampling in current observations, it exemplifies how much the observational constraints on ICM magnetic fields depend on the adoption of specific three-dimensional models. RM studies must rely on discrete spatial sampling of the cluster in the plane of the sky, which makes the magnetic field strength reconstruction sensitive to the shape of the power spectrum at a given position in that plane.
%TOLTO IN QUANTO RIPETIZIONE The most common procedure to produce mock RM maps relies on the assumption of a density profile for the ICM, that the magnetic field strength scales as a power law of the ICM density and that the fluctuations of the magnetic field are random realizations according to a power spectrum  \citep{Murgia04,Vacca10,Bonafede13,Stuardi21,Alonso26}.
While the situation is expected to greatly improve with the deployment of the Square Kilometre Array\footnote{\hyperlink{https://www.skao.int/en}{https://www.skao.int/en}} \citep[SKA; e.g.][for recent reviews]{Vacca26,OSullivan26,loi26_ska}, at present, the number of detected polarized radio sources is generally insufficient to uniformly cover the entire cluster extent in the plane of the sky. 
%RM observations trace the mean magnetic field component along the LOS, but they do not directly probe the magnetic-field dispersion along the LOS. 
%This limits the possibility of deriving the magnetic power spectrum using RM maps alone, in particular at high wavenumbers.
This makes the adoption of theoretically motivated magnetic field models particularly relevant in view of current and forthcoming facilities, such as SKA, which are expected to substantially extend the coverage of RM data out to larger cluster radii, and therefore probe magnetism in regions where simplistic assumptions of symmetric and homogeneous ICM conditions may fail. 

From the theoretical viewpoint, the magnetic field in the ICM is thought to arise from the amplification of extremely weak seed fields -- either generated in the early Universe or injected by astrophysical sources such as active galactic nuclei, and subsequently amplified during cluster assembly in two ways \citep[][for a review]{Donnert18}. 
1) The cluster assembly compresses the plasma, thereby amplifying the magnetic field of the ICM, but hardly up to the observed $\sim \muG$ values across the full extent of galaxy clusters. 
%This process may be the main reason for the dependence of the magnetic field strength on ICM density as found in both real and simulated clusters\footnote{The magnitude of the magnetic field, amplified only by plasma compression, scales with the ICM density as a power law with exponent 2/3. Similar exponents are found in the spherically-averaged profile of the magnetic field strength in clusters formed in cosmological simulations.}.
%Equally importantly, 
2) It injects turbulent motions in the ICM. 
These velocity fluctuations are typically of the order of a few hundred $\kms$ in the ICM and have been detected in the inner regions of galaxy clusters with the currently operating X-ray spectrometer XRISM\footnote{\hyperref[https://www.xrism.jaxa.jp/en/]{https://www.xrism.jaxa.jp/en/}} \citep[e.g.,][]{XRISM_A2029_1, Bartalesi26, Zhang26}. 
These velocity fluctuations amplify the ICM magnetic field by stretching and folding magnetic field lines and promoting a small-scale (turbulent or fluctuation) dynamo, a mechanism that remains not fully understood even in controlled numerical experiments \citep[e.g.][]{Schekochihin02,Schekochihin04,Rincon19}. 
Establishing how it depends on the assembly history, as well as clarifying the dynamo process itself, remains a key open problem.
%In the framework of the fluctuation dynamo, which is the theory describing the interplay at disparate scales between velocity and magnetic fluctuations, the amplification of the magnetic field strongly depends on the Reynolds number \citep[e.g.,][]{Schober15, Donnert18}.

Cosmological magneto-hydrodynamical (MHD) simulations are routinely used to model the cluster assembly.
%However, the finite spatial resolution limits the achievable Reynolds numbers in these simulations to values orders of magnitude lower than those predicted for the ICM, making the fluctuation dynamo in cosmological simulations potentially less efficient than in real systems \citep{Donnert18}.
 \citet[hereafter \citetalias{Vazza18_magne}]{Vazza18_magne} reported evidence of spatially-resolved small-scale dynamo amplifications in simulated clusters, leading to a realistic magnetic field when compared to RM observations. 
Analyzing the same set of simulations and limiting to the innermost $\approx 2\, \rm Mpc^3$ cluster regions, \citet[hereafter \citetalias{Dominguez19}]{Dominguez19} have further derived a functional form that appropriately describes the power spectrum of the magnetic field resulting from fluctuation dynamo amplification. 

Our work aims to improve theoretical modeling and understanding of the magnetic field power spectrum in galaxy clusters.
Specifically, we provide a more appropriate functional form to represent the power spectrum of the magnetic field at different stages of small-scale dynamo in the simulated ICM, based on one Coma-like cluster simulation. 
We then parameterize the dependence of this functional form's parameters on local, observationally accessible ICM properties. 

%...secondo me non rilevante qui 
%The choice of analyzing a Coma-like cluster has the following motivations.
%Providing the tightest constraints thus far on the magnetic field, Coma is considered the natural testbed for each theoretical study on the ICM magnetic field. 
%Having disturbed morphology associated with recent or ongoing mass accretion\tb{Per me, mass accretion includes merging activity. Poi scegliete voi se preferite sostituire mass accretion with merging activity.}, clusters similar to Coma are promising candidates for hosting, at least, a radio relic; as with evident surface-brightness fluctuations interpreted as signatures of ICM turbulence\tb{Volete parlare subito di turbulence invece di surface-brightness fluctuations?}, these clusters are expected to exhibit a high RM signal due to the magnetic field amplification via fluctuation dynamo. 
The paper is organized as follows.
Section \ref{sec:simulation} introduces the simulation data and the procedure to derive power spectra.
Sect. \ref{sec:model} describes the model for the power spectrum of the magnetic field that we propose in this paper.
Sect. \ref{sec:results} presents and discusses the fitting to the magnetic power spectrum, while Sect. \ref{sec:correlation} studies the dependence of the parameter of the latter on the variations of the ICM conditions across the cluster.
Sect. \ref{sec:conclusions} concludes.
%In this work, we assume a flat $\Lambda$ cold dark matter cosmological model with the present-day matter density parameter $\Omega_{\mathrm{m},0}=0.3$ and Hubble constant $H_0=70\,\kms \mathrm{Mpc^{-1}}$.  
To evaluate the statistical significance of two-parameter correlation, this work widely uses Spearman's correlation coefficient, $\rs$.
Throughout this work, we define modest and strong correlations for $0.3 \leq |\rs| < 0.6$ and $|\rs| \geq 0.6$, respectively.

%\footnote{The turbolence-in-a-box simulations often assume a homogeneous, isotropic and incompressible plasma in a periodic box. The energy injection occurs through a random, non-helical, time-uncorrelated external forcing, applied to the largest scales in the box.}.
%Tightly connected to this 

\section{Data from a cosmological zoom-in simulation}
\label{sec:simulation}

%In Sect. \ref{sec:setup}, we summarize the setup and characteristics of the cosmological simulation by \citetalias{Vazza18_magne}, while in Sect. \ref{sec:fourier} we describe the procedure to analyze the snapshot of the selected Coma-like cluster and to obtain power spectra from it.

\subsection{Simulation setup}
\label{sec:setup}

We use the publicly available\footnote{\hyperref[https://cosmosimfrazza.eu/]{https://cosmosimfrazza.eu/}} $z=0$ snapshot of the cluster E18B simulated by \citetalias{Vazza18_magne} using the Eulerian {\tt Enzo}\footnote{\hyperref[https://enzo-project.org/]{https://enzo-project.org/}} code, with the Dedner formulation of the magneto-hydrodynamic (MHD) equations and an adaptive mesh refinement (AMR) algorithm with eight levels. 
The simulation includes only cosmic expansion, gravity and ideal MHD. 
The cluster has a virial mass of $\sim 10^{15}$ M$_{\odot}$ and it experienced its last merger at $z \approx 0.5$.  
E18B was selected based on its similarity with the cluster Coma, which, being intercepted along the LOS by some background radio galaxies and having one radio relic in the outskirts, provides the tightest constraints to date on the magnetic field of the ICM.
Indeed, \citetalias{Vazza18_magne} found a good match of the RM properties of E18B to those observed in Coma \citep{Bonafede10, Bonafede13} and the same system was used to help the interpretation of the turbulent properties of Coma, as observed using XRISM \citep[][]{va26xrism}.

At the initial redshift of the simulation ($z = 30$), \citetalias{Vazza18_magne} assumed a uniform primordial magnetic field, with a strength of 0.1 nG (comoving) in each Cartesian direction.

\begin{figure}
   \centering
   \includegraphics[width=0.49\textwidth]{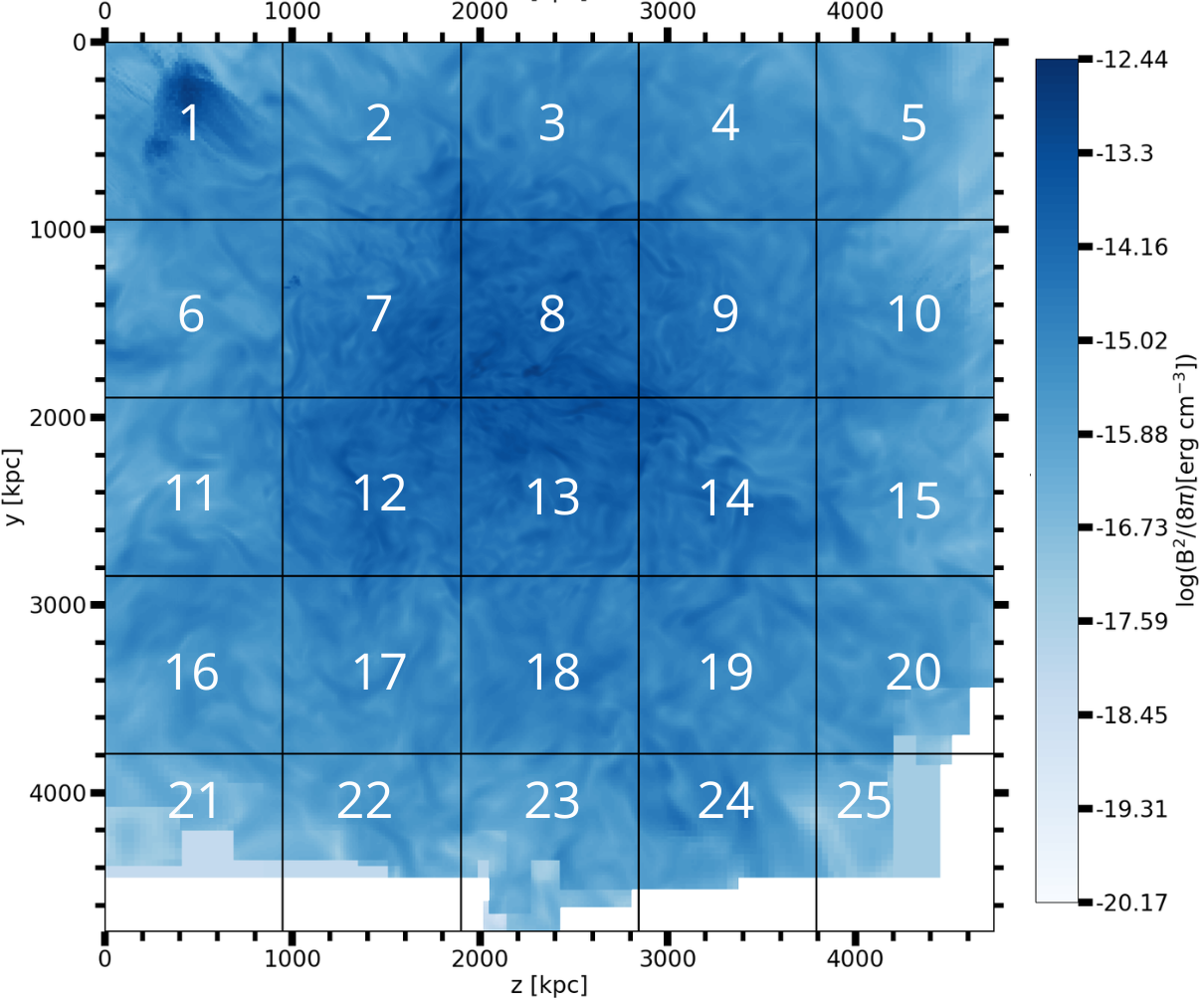}
    \caption{Volume-weighted projected magnetic energy density map for a central 4.74$^2$ Mpc$^2$ slice of our cluster at $z=0$, with thickness equal to $\approx 950$ kpc corresponding to the subbox side length $L$ (see Sect. \ref{sec:fourier}). Pixels contaminated by the projected contribution from regions only sampled with resolution coarser than 15.8 kpc have been set to 0, to signal that in our spectral analysis we try to minimize or exclude their contribution (see Sect. \ref{sec:posterior} for details). 
    The black grid defines the projected contour of the cells defined in Sect. \ref{sec:fourier}.
    The numbers indicated for each subbox are the same used in Figs. \ref{fig:map_15} - \ref{fig:map_30}.}
    \label{fig:mapB}
\end{figure}

Here we analyze the $\approx 25$ Mpc$^3$ comoving cubic volume at $z=0$, approximately centered on the cluster center of mass.  
The finest spatial resolution of the simulation is $\approx$ 3.95 comoving kpc; the corresponding effective Reynolds number \citep[see Sect. 4.4.1 of][for the definition]{Donnert18} in the cluster outskirts is $R_e\lesssim 500$ \citepalias[while somewhat higher values can be  estimated in the central 2 Mpc$^3$ cubic volume; see Sect. 4.1.1 of][for details]{Vazza18_magne}.
Where the simulation does not reach the finest resolution, the quantities stored in the $z=0$ snapshot are interpolated onto a uniform, linearly spaced grid with a cell size of 3.95 kpc during post-processing.
%\cite{Vazza18_magne} perform six simulations of E18B progressively refining the spatial resolution of the simulation by factors of 2: the spherically-averaged profile of the magnetic field strength only at radii larger than 700 kpc converges in the two finest runs. 
%\tb{Secondo voi, devo fare un altro riferimento al fatto che Franco ha mostrato evidenza di small-scale dynamo in questa simulazione.}
Figure \ref{fig:mapB} shows the projected magnetic energy density for a central slice of our cluster at $z=0$ (see Sect. \ref{sec:fourier}, for definition), where we also indicate the number of each subbox used for the evaluation of magnetic spectra in Figs. \ref{fig:map_15} and \ref{fig:map_30}. 
The maps of projected ICM density, kinetic energy density, and magnetic energy density for our cluster slices are available at \myzenodo.

\subsection{Magnetic and velocity power spectra}
\label{sec:fourier}

%The three Cartesian directions $x$, $y$ and $z$ in this snapshot are defined to be parallel to the orthogonal sides of this volume.
The data for each Cartesian component of the velocity and magnetic field of the simulated cluster at $z=0$ are on a 3D uniformly (linearly) spaced grid, covering a central cubic box of side length $\approx 5$ Mpc, approximately centered on the total mass.
To minimize the spurious contribution from cells that are not refined up to the highest AMR level, which are found in some cases at the boundary of this central volume, we conservatively further restricted the volume for our spectral analysis to a slightly smaller 4.74$^3$ Mpc$^3$ cubic box, which we divided into 125 identical subboxes, each with side length $L\approx$ 950 kpc (corresponding to $N = 240$ cells of size length of 3.95 kpc).
This choice of $L$ is empirically motivated by the trade-off between covering the reduced volume with the largest possible number of subboxes and having a statistically representative number of data points per subbox. 
Moreover, the choice of $L\approx$ 950 kpc allows us to monitor the spectral properties of the magnetic field on a scale comparable to the typical linear extent of radio relics or radio halos \ \citep[e.g.,][]{vanWeeren19}.
The projected contours of the cells in the central slice of our cluster are shown in Fig. \ref{fig:mapB}.

We now illustrate the procedure for deriving, from the simulation, the power spectrum data for the $h$-component of the vector field $\vec w$, $w_h(x, y, z)$, where $h = \{x, y, z\}$. 
We work in the 3D Fourier space, using the convention that each discrete coordinate, $k_h$, takes the ordered values $\{0, 1, ..., N / 2 -1, -N/2, ..., -1\}$.
In this formalism, the value of $w_h$ evaluated at any point of the subbox, indicated with $(j_x, j_y, j_z)$, $w_{h, j_x, j_y, j_z}$, is a random realization of $w_h$ according to the power spectrum, which we want to derive here.
Under the assumption that $w_{h, j_x, j_y, j_z}$ is a periodic sequence of period $(N, N, N)$, the discrete fourier transform of $w_{h, j_x, j_y, j_z}$ is
\begin{equation}
    \tilde w_{h,k_x, k_y, k_z} = \sum_{j_x=0}^{N-1} \sum_{j_y=0}^{N-1} \sum_{j_z=0}^{N-1} w_{h, j_x, j_y, j_z} \exp \left[-\frac{2\pi i}{N} \left(k_x j_x + k_yj_y + k_zj_z\right) \right].
    \label{eq:fourier}
\end{equation}

Because $\tilde w_{h,-k_x,-k_y,-k_z}$ is the complex conjugate of $\tilde w_{h,k_x,k_y,k_z}$, the Fourier modes are not independent. We therefore restrict the analysis to modes with arbitrary $k_x$, $k_y$, and $k_z\geq 0$, thereby eliminating mode redundancy.
In the 3D Fourier space, we define $k_d = \sqrt{k_x^2 + k_y^2 + k_z^2}$ as the distance of the generic point $(k_x, k_y, k_z)$ from the origin.
The power spectrum of $w_{h}$ is defined at any $r$ as $P_{\mathrm{w}, h,r} = \langle |\tilde w_{h}|^2 \rangle_r$, where $\langle...\rangle_r$ indicates the average over all the modes with $r - 1/2 \leq k_d < r+1/2$ and $r = \{0,1, ..., N/2 - 1, -N/2, ..., -1\}$. 
We estimate the error on $P_{\mathrm{w}, h,r}$ at a given $r$ among all $|\tilde w_{h, k_x, k_y, k_z}|^2$ lying in this $k_d$ bin as the standard error on the mean.
Thus we obtain the 3D power spectrum of the vector field $\vec w$ as $P_{\mathrm{w},r} = \sum_{h = 1}^{3} P_{\mathrm{w}, h, r}$ and the error on $P_{\mathrm{w}, r}$ as the sum in quadrature of the standard errors on the mean of each $h$-component at fixed $k = r / L$.
As reported in Sect. \ref{sec:setup}, in low density subboxes, where the reconstructed field $\vec w$ results from the interpolation of a coarser resolution mesh or is artificially set to zero during post-processing, we do not have exactly $N^3$ independent and nonnull cell values, which can lead to numerical (unphysical) spikes in power spectrum (see Sect. \ref{sec:results}, for details). %and is not accounted for in our error budget.

%The time sampling of the simulated cluster evolution is obtained with a series of saved snapshots distant by many timesteps in the simulation. 
%Given that the vector field $\vec w$, under stimulation by the ongoing cluster assembly, evolves more rapidly than this coarse time sampling, our error budget cannot account for the statistical errors associated with the random behavior of $\vec w$ at a fixed comoving position at different timesteps of the simulation.

\section{A model for the magnetic power spectrum}
\label{sec:model}

This Section describes the functional form used to fit the magnetic power spectra in different subboxes of our snapshot, following the original work by \citetalias{Dominguez19}.
Our goal is to find a functional form sufficiently flexible to reproduce the data in each subbox, with parameters that reflect the diversity of ICM conditions across different locations in the cluster volume.

We model the shape of the magnetic power spectrum using the functional form introduced by \citetalias{Dominguez19}, which is inspired by the analytical work of \citet{Kulsrud92}. 
The latter solved the kinetic equation describing the transfer of energy across magnetic modes through interactions with turbulent velocity modes. 
This equation is valid under the assumptions that the evolution of the magnetic field is slower than the inverse decorrelation time of the turbulent modes and that the turbulent cascade is not significantly modified by dynamo amplification. 
The phenomenological parameterization proposed by \citetalias{Dominguez19} (see below)
%This parametrization (hereafter the DF19 model) 
was shown to well reproduce the 3D magnetic power spectrum of the ICM within the innermost 2$^3$ Mpc$^3$ regions of several clusters, included the one we analyze here.
However, in this work, we attempt to apply the same approach to parametrize the shape of the magnetic power spectra also in more peripheral regions, where the departures from the initial assumptions of \citet{Kulsrud92} are even more pronounced, due to the presence of intermittent forcing, anisotropic bulk motions and the mixing of magnetic patches with different amplification histories in cluster outskirts.  

%At small wavenumbers, the solution contains the "Kazantsev" solution \citep{Kazantsev68}, i.e. an analytical model for the fluctuation dynamo under idealized conditions for the plasma, such as a Dirac delta correlated velocity field, incompressibility and isotropy, which predicts a magnetic power spectrum characterized by a $k^{3/2}$ scaling at small wavenumbers, a peak at intermediate $k$ and a very rapid decline towards large $k$.
%This functional form (hereafter, the DF19 model) has been successfully applied by \cite{Dominguez19} to reproduce the power spectrum of the magnetic field of the ICM for the innermost $\sim 2^3$ Mpc$^3$ regions of the snapshots of several massive clusters, including the one we analyze here.  
%In what follows, we shall refer to this model as to the DF19 model, for short. 

Numerical simulations typically show that, where the dynamo amplification is not developed (either for physical or numerical reasons), the magnetic field structure retains memory of large-scale stirring motions before structure formation, which can be approximated by a Kolmogorov-like ($\propto k^{-5/3}$ ) power spectrum, i.e., exactly following the velocity power spectrum \citep[e.g.][]{Cho09, Vazza14,Vazza18_magne,Seta20,Sur24}.

Therefore, we extend the previous DF19 model by including an additional Kolmogorov-like term, in order to account for the contribution from magnetic field structures in our subboxes (especially the most peripheral ones) that are not fully in a dynamo amplification stage.
The model for the power spectrum of the magnetic field we propose in this work is 
\begin{equation}
    \EM(k) = A\tilde k^{3/2} \left\{1- \erf\left[B \ln\left(\lC  k\right)\right]\right\} + D \exp \left(\frac{-1}{\lF k}\right) \tilde k^{-5/3},
    \label{eq:model}
\end{equation}
where the last addend is our addition to the original DF19 model \citepalias[][]{Dominguez19}, while $\erf$ is the error function. 
We define $\tilde{k} = kL$, where $k$ and $L$ are the wavenumber and the subbox side length, as defined in Sect. \ref{sec:fourier}, so that $\tilde k$ is dimensionless. 
$A$ and $D$ are the normalizations of the asymptotic behaviors of DF19 and Kolmogorov-like terms, $2A\tilde k^{3/2}$ for $k\ll \lC^{-1}$ and $D\tilde k^{-5/3}$ for $k\gg \lF^{-1}$, respectively.
$B$ and $\lC$\footnote{Our parameter $\lC$ is equal to $1/C$ of Eq. (4) in \cite{Dominguez19}.} indicate the width and scale of the peak of the DF19 model\footnote{\citet{Kulsrud92} predicted the 3D magnetic power spectrum as a function of the time and wavenumber. 
At every time, it scales as $k^{3/2}$ at small wavenumbers, consistent with the Kazantsev solution \citep{Kazantsev68}, and exhibits a steep decline at large $k$. 
The transition between these two regimes occurs at a characteristic wavenumber corresponding to the spatial scale containing the largest fraction of the magnetic energy and to $1 / \lC$ in Eq. (\ref{eq:model}), which \citetalias{Dominguez19} treated as a free parameter.
\citetalias{Dominguez19} replaced the time evolution of the \citet{Kulsrud92} power spectrum with the two free parameters, $A$ and $B$.}, respectively, whereas $\lF$ is the scale of the exponential cut-off of the Kolmogorov-like term at relatively small wavenumbers.
The figures in Appendix \ref{sec:flexibility} show how such a model for the magnetic power spectrum varies with the parameters $B$, $\lC$, $\lF$ and the ratio $D/A$.

%\begin{equation}
    %\label{eq:power}
    
%\end{equation}

\section{Fitting the data of the magnetic power spectrum}
\label{sec:results}
In this Section, we analyze the magnetic power spectrum of the ICM in individual subboxes of the simulation snapshot presented in Fig. \ref{fig:mapB} (see Sect. \ref{sec:setup}, for details).
For the interpretation of power spectra, it is important to know the Nyquist wavenumber, defined according to the  
Nyquist-Shannon sampling theorem, as $\kny = \Nindep / 2L$, where $\Nindep$ is the number of independent points along each Cartesian direction within a given subbox and $L$ is the subbox side length. 
In practice, $\kny$ represents the largest wavenumber for which the power spectrum is accurately obtained.
While the estimate of $\kny$ is trivial in a fix grid simulation, it gets more complex for simulations with several resolution levels, like in the case of the AMR run we have here. 
When the entire simulated volume has a uniform resolution of 3.95 kpc, $\Nindep = N$, which is the number of grid points along each direction, as defined in Sect. \ref{sec:fourier}.
When these $N^3$ points of each subbox are obtained from the interpolation of nested AMR levels over a uniform grid, the analytical evaluation of $\kny$ is impractical, and the choice $\Nindep = N$ leads to a too large $\kny$ and, as a consequence, includes numerical spikes in the power spectra \citep[e.g.,][]{Martin22}.
We therefore evaluate $\kny$ considering two effective spatial resolutions of the simulation.
First, we conservatively assume the simulation was run with a uniform spatial resolution of 32 comoving kpc, such that $\Nindep = N/8$ and $\kny L = 15$ (corresponding to a fluctuation scale of approximately 60 kpc).
Under such an assumption, in Sect. \ref{sec:statistics}, we analyze the magnetic power spectra of the ICM in the individual subboxes defined in Sect. \ref{sec:fourier}, using the original DF19 model.
This choice is extremely conservative because \citetalias{Vazza18_magne} found that most of the simulated volume reaches a spatial resolution of 16 kpc or better.

For nearby clusters, the angular resolution of radio interferometers corresponds to scales of only a few kpc. 
Restricting observational analyses of the RM maps to magnetic fluctuations on scales larger than 32 kpc, or extrapolating our model-inferred results up to $ \kny L = 15$ down to the scales probed by the observations, potentially limits the accuracy of the constraints on the magnetic field strength profile. 
To gain insight into the behavior of the 3D magnetic power spectrum at higher wavenumbers, we also explore less conservative assumptions for $\kny$ in Sect. \ref{sec:posterior}, where we extend our analysis to wavenumbers for which the magnetic power spectrum may also be affected by the numerical artifacts discussed above.
Specifically, in Sect. \ref{sec:posterior}, we study the magnetic power spectra of the ICM in the same individual subboxes as in Sect. \ref{sec:statistics}, but assuming $\Nindep = N / 4$ (corresponding to a uniform spatial resolution of 16 kpc) and $\kny L = 30$ (corresponding to a fluctuation scale of approximately 30 kpc).

%Sect. \ref{sec:statistics} describes the statistical method used to fit Eq. \ref{eq:model} to the data of the power spectrum of the magnetic field in the subboxes in which we divide the E18B $z=0$ snapshot, as detailed in Sect. \ref{sec:fourier}.
%Sect. \ref{sec:posterior} presents and discusses the results of this fitting.

\begin{table*}
\caption[]{Prior and posterior values of the MCMC parameters.}
\centering
\begin{tabular}{l c ccc ccc c}
\hline
\noalign{\smallskip}
Parameter & prior & \multicolumn{3}{c}{$D=0$ posterior} & \multicolumn{3}{c}{$D\neq0$ posterior} & $\mathcal{R}$ \\
 &  & 5th & 50th & 95th & 5th & 50th & 95th & \\
\noalign{\smallskip}
\hline
\noalign{\smallskip}
$\log (A /[10^{-9}\mathrm{erg \, cm^{-3}}] )$ & [-2.8, 1.85] & -2.05 & -0.35 & 0.48 & -2.31 & -0.55 & -0.36 & 0.06 \\
$B$ & [0.3, 2.2] & 1.12 & 1.31 & 1.51 & 1.20 & 1.50 & 1.78  & 0.15\\
$L / \lC$ & [2, $\kny L$] & 2.02 & 3.34 & 5.81 & 2.34 & 3.91 & 6.45  & 0.16\\
$\log (E/A)$ & [-2, 2] & -- & -- & -- & -0.91 & 1.05 & 1.55  & 0.18\\
$\log (L / \lF)$ & [-1, 1.5] & -- & -- & -- & -0.59 & 0.32 & 1.34 & 0.30\\
\noalign{\smallskip}
\hline
\end{tabular}

\tablefoot{First column: parameter names. Second column: prior ranges. Third and fourth columns: percentiles of the marginal total posteriors for the dynamo-like ($D=0$) and extended ($D\neq 0$) models, respectively. Fifth column: ratio, $\mathcal R$, between the median 1$\sigma$ relative uncertainty (see text) of individual subboxes in the parameter estimate and the corresponding uncertainty from the marginal total posteriors for the $D\neq 0$ models. }
\label{tab:fitting}
\end{table*}

\subsection{Results for the dynamo-only model $(D=0)$}
\label{sec:statistics}

\begin{figure*}
   \centering
   \includegraphics[width=1\textwidth]{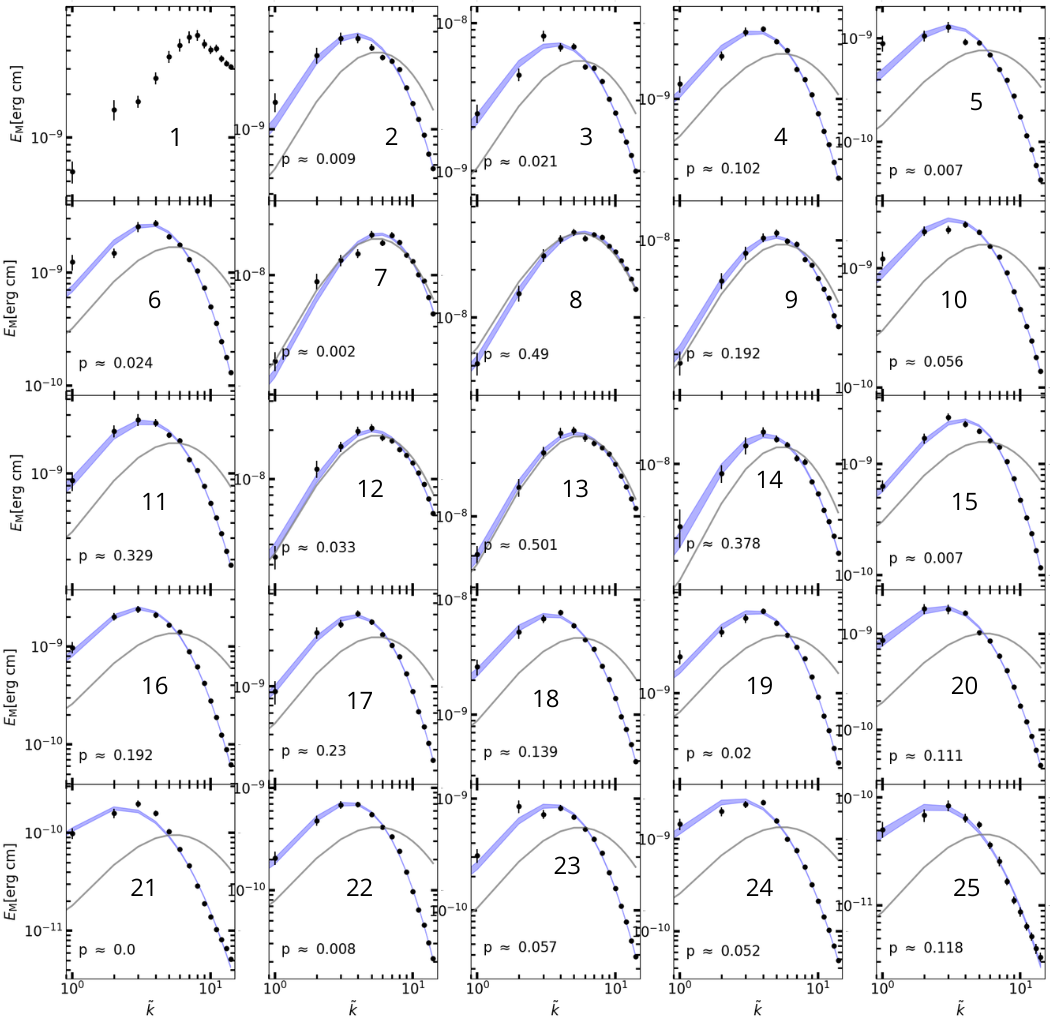}
    \caption{3D power spectra of the magnetic field in the $\tilde k$ ranges 0 -- 15 for the E18B central slice, with thickness equal to the subbox side length ($L\approx 950$ kpc).
    The black points with error bars are the magnetic power spectrum data, whereas the blue regions are the 16th-84th percentile interval of the DF19 model.
    For comparison, we overplot the best-fit magnetic power spectrum of \citetalias{Dominguez19} (gray lines), renormalised to match the total magnetic energy in the subbox (see text for details). The numbers from 1 to 25 refer to the same subboxes as Fig. \ref{fig:mapB}.
   }
    \label{fig:map_15}
\end{figure*}

Following Section~\ref{eq:fourier}, in each subbox of our cluster at $z=0$ for each wavenumber $k$ we derive the 3D power spectrum of the magnetic field, $P_{\mathrm{M}}(k)$. % with $w=M$.
Given that the energy of the magnetic fluctuations within the subbox is $D_\mathrm{B}=4\pi \int dk k^2 P_{\mathrm{M}}(k) $, it is convenient to define the quantity $\EMk = k^2 P_\mathrm{M}(k)$, which represents the energy per wavenumber.
In Fig. \ref{fig:map_15}, we compare the data of the 3D magnetic power spectrum of individual subboxes of our central slice (see Sect. \ref{sec:fourier}) with the best-fit model given by \citetalias{Dominguez19} and based on the analysis of the innermost (2 Mpc)$^3$ subbox of the $z=0$ snapshot of the same cluster used here (in \myzenodo \ we report the same figures as \ref{fig:map_15}, but for all the slice in which we divide our cluster in Sect.~\ref{sec:fourier}). 
We show this in all subboxes to better highlight how the \citetalias{Dominguez19} best-fit model works extremely well to describe the magnetic power spectrum in the cluster center, while it progressively loses its capability of reproducing the magnetic spectra -- as expected -- due to a loss of efficiency of the small-scale dynamo amplification moving towards increasingly more peripheral regions.
For this comparison, in all subboxes we set the normalization parameter of the best-fit formula of \citetalias{Dominguez19}, $A$ in Eq. (\ref{eq:model}) with $D=0$, so that the integral over the power spectrum of this model gives the same total energy as after integrating our $\EMk$ spectrum, within the same range of wavenumbers. 
We have left the remaining parameters, instead, fixed to the best-fit values listed in Table 1 of \citetalias{Dominguez19}\footnote{The best-fit values of \citetalias{Dominguez19} in the convection used by Eq. (\ref{eq:model}) are $B =1.11$ and $\lC = 200$ kpc.}.
%The best-fit model of \citetalias{Dominguez19} closely matches our data only in the central subbox in the left panel of Fig. \ref{fig:map}, which encompasses the cluster mass center.
The agreement between our data and the best-fit model of \citetalias{Dominguez19} progressively deteriorates towards the cluster outskirts, i.e., when moving from the central subboxes to the edges of the map in Fig. \ref{fig:mapB}. 

%We start investigating this tendency, fitting the Kulsrud model, introduced in the study of the magnetic field in galaxy clusters by \citetalias{Dominguez19}, to individual subboxes.

%Given that most of the simulated volume is refined up to the third AMR level\tb{Si capisce che il primo AMR level è quello con la risoluzione più fine?}, corresponding to a spatial resolution of $\approx$ 16 comoving kpc \citep[see Sect. 2 of][]{Vazza18_magne}, we conservatively consider power spectra up to the Nyquist frequency of the third AMR level mesh, corresponding to $kL = 30$. 
%However, given that some regions are simulated at coarser resolution than 16 kpc, the interpolation of raw data onto a uniform mesh of $\approx$ 4 kpc (see Sect. \ref{sec:setup}) is bound to introduce some spurious numerical noise and spikes in the magnetic energy spectra \citep[e.g.,][]{Martin22}. 
%This problem is only severe in the case of the most peripheral sub-boxes of our collection, where the drop in density is followed by a decrease in the spatial resolution. 
%In addition, we exclude from our analysis the subbox ... whose map of the magnetic ene, it is standard to work withrgy shows the presence of a clump, with a size of 300 kpc and a magnetic energy more than one order of magnitude higher than in the neighbors, because it cannot be described within the formalism adopted for this work in Sect. \ref{sec:fourier}.

Given that our analysis interprets the shape of the 3D power spectrum of the magnetic field based on dynamo-like models, we first define a subbox collection by simply visually inspecting the magnetic power spectra of individual subboxes over the $\tilde k =kL$ range 1 -- 15 and rejecting those with spectra manifestly inconsistent with the dynamo model. 
Specifically, from our collection, we exclude subboxes in which the data of the 3D magnetic power spectrum do not exhibit a peak at intermediate $k$ and a decline towards both smaller and larger $k$ (see Sect.~\ref{sec:model}). 
Only four subboxes are removed in this way (those with only the data in Fig. \ref{fig:map_15} and in \myzenodo). 
Notably, all of them lie at the edges of the simulated volume (see Fig.\ref{fig:map_15}), where the simulation has the coarsest spatial resolution, consistent with the fact that we have an inefficient (or just absent) fluctuation dynamo there.

Next, by fixing $D = 0$, we fit the model for the 3D magnetic power spectrum in Eq. (\ref{eq:model}) to the corresponding data in the $\tilde k $ range (1 -- 15), for each subbox individually, via a Markov chain Monte Carlo (MCMC) algorithm that maximizes the natural logarithm of the posterior distribution values.
The parameter vector of the posterior, $\vec \theta$, is composed by $\log (A /[10^{-9}\mathrm{erg \, cm^{-3}}])$, $B$ and $L / \lC$. 
In our analysis, we assume uniform priors on these parameters, with lower and upper bounds identical for all the subboxes, reported in Table \ref{tab:fitting}.
We assume that the datum of the 3D magnetic power spectrum at a given $k$ is randomly generated from a normal distribution with median and standard deviation equal to the natural logarithm of the model values and the relative error, respectively, as defined in Sect. \ref{sec:fourier}.
As posterior sampling of the MCMC run for each subbox that we analyze here, we randomly select 1000 $\vec \theta$ from the MCMC sampling\footnote{We run the MCMC with 50 walkers for 5000 iterations. We empirically notice that the ln-posterior distribution becomes stationary, as the iteration number increases, after 1200 iterations. As a rule for each subbox, we conservatively take the end of the burn-in phase after 2500 iterations. %Given that we empirically notice drastic variations in the ln-posterior value occurring at least after 20 iterations across the stationary phase, we conservatively assume uncorrelated chains after 20 iterations. We obtain our posterior sampling by thinning the remaining 500 iterations by this auto-correlation length.
We thin the remaining sampling by 50 iterations.}.

%The relative uncertainty on the data points typically varies by three orders of magnitude from $kL = 1$ to $kL=30$, with the larger modes being more uncertain primarily due to reduced spatial sampling in the evaluation of the power spectrum in Sect. \ref{sec:fourier}\tb{Per voi, ha senso dire che visual comparisons sono rischiose?}. 
Following \cite{Gelman13}, we evaluate the appropriateness of the fitting for each subbox through the Bayesian p-value.
Taking into account the uncertainties in both the data and the model parameters, the Bayesian p-value quantifies the probability that the model in Eq. (\ref{eq:model}) generates mock data, according to the posterior distribution, that are as extreme as or more extreme than the actual data.
Specifically, for the actual dataset of each subbox, we evaluate 1000 $\chi$-squared values, each with respect to one of the 1000 $\vec \theta$ sampled from the posterior.
At each $\tilde k$ in the range (1 -- 15), we generate a mock dataset for each parameter vector, $\vec \theta$, in the posterior sampling, by randomly extracting a mock data point from a normal probability distribution function with median and standard deviation equal to the corresponding natural logarithmic of the value in Eq. (\ref{eq:model}) and to the corresponding relative error in the data, respectively. 
We then evaluate the chi-squared statistic for this mock dataset with respect to the power spectrum model used to generate it.
The Bayesian p-value is the fraction of the posterior samples with a chi-squared statistic for the actual dataset smaller than that of the corresponding mock dataset. % $p = Pr(\chi^2_\mathrm{mocks} \geq \chi^2_\mathrm{data})$.
We report this p-value in each subbox in the upper left of each panel in Fig. \ref{fig:map_15}.
%Though studying the cause of a low Bayesian p-value goes beyond the scope of this work, it is worth mentioning that a low Bayesian p-value can result not only from the use of an inappropriate model, but also from the inappropriateness of our periodicity assumption (see Sect. \ref{eq:fourier}, for details), error underestimation (see Sect. \ref{sec:fourier}, for details), overdispersion of the data around the model values or data already contaminated by numerical effects. 
We assume that, in a given subbox, the posterior distribution of our model provides an appropriate description of the data for the magnetic power spectrum, only if the corresponding Bayesian p-value is greater than the threshold $t$, with $0 \leq t<1$.
In this way, $t$ quantifies the probability below which the data under consideration are too unlikely to be generated from the corresponding model posterior.
Given that random realizations of a model produce a distribution of Bayesian p-values, but we select the appropriate fits based on a single value $t$, this selection can potentially produce false positives or false negatives.
To mitigate the impact of the latter on our results, we perform two parallel analyses using $t = 0.01$ and $t = 0.05$ and compare the results.

Figure~\ref{fig:map_15} also shows the comparison of the model for the 3D power spectrum of the magnetic field of the ICM obtained from our posterior sampling in each subbox with the corresponding data as derived in Sect. \ref{sec:fourier} from the E18B snapshot at $z=0$ over the $\tilde k$ range (0 -- 15).
As $\approx$ 44\% and 70\% of subboxes have a Bayesian p-value above $t = 0.05$ and $t=0.01$, respectively, the DF19 model (Eq. (\ref{eq:model}) with $D=0$) well reproduces the data of the 3D power spectrum of the magnetic field of the ICM individually for a large collection of our subboxes.
This provides evidence that a magnetic field primarily arises from a fluctuation dynamo throughout most of the cluster volume.
In Tab. \ref{tab:fitting}, we report some percentiles obtained by summing the posterior sampling of all the individual subboxes with a Bayesian p-value higher than $t=0.01$.
However, the Bayesian p-value, reported in each subbox of Fig. \ref{fig:map_15}, tends to be, on average, lower moving towards the outskirts.
This is particularly evident in slices along the edges of the simulated volume (Figs.~5 and 29 in \myzenodo), where many subboxes have Bayesian p-values of 0.
This indicates that the DF19 model, inspired by dynamo results, performs better in the central parts of the cluster (that was the target of the DF19 work) than in the outskirts, %is not flexible enough to entirely reproduce the data of the 3D magnetic power spectrum across the full cluster volume, 
and specifically in all regions where the fluctuation dynamo does not appear to dominate the magnetic field distribution, and/or there are more magnetic field components at play. 
This can be due either to quenching of the dynamo process because of a Reynolds number artificially limited by finite spatial resolution, or to physically unsuitable conditions for the fluctuation dynamo, such as the predominance of compressive and supersonic motions. 
We thus apply in Sect.~\ref{sec:posterior} the extended ($D \neq 0$) model in Eq.~(\ref{eq:model}), which includes a model component for the 3D magnetic power spectrum of the ICM in the presence of an inefficient dynamo.

\begin{figure*}
   \centering
   \includegraphics[width=1\textwidth]{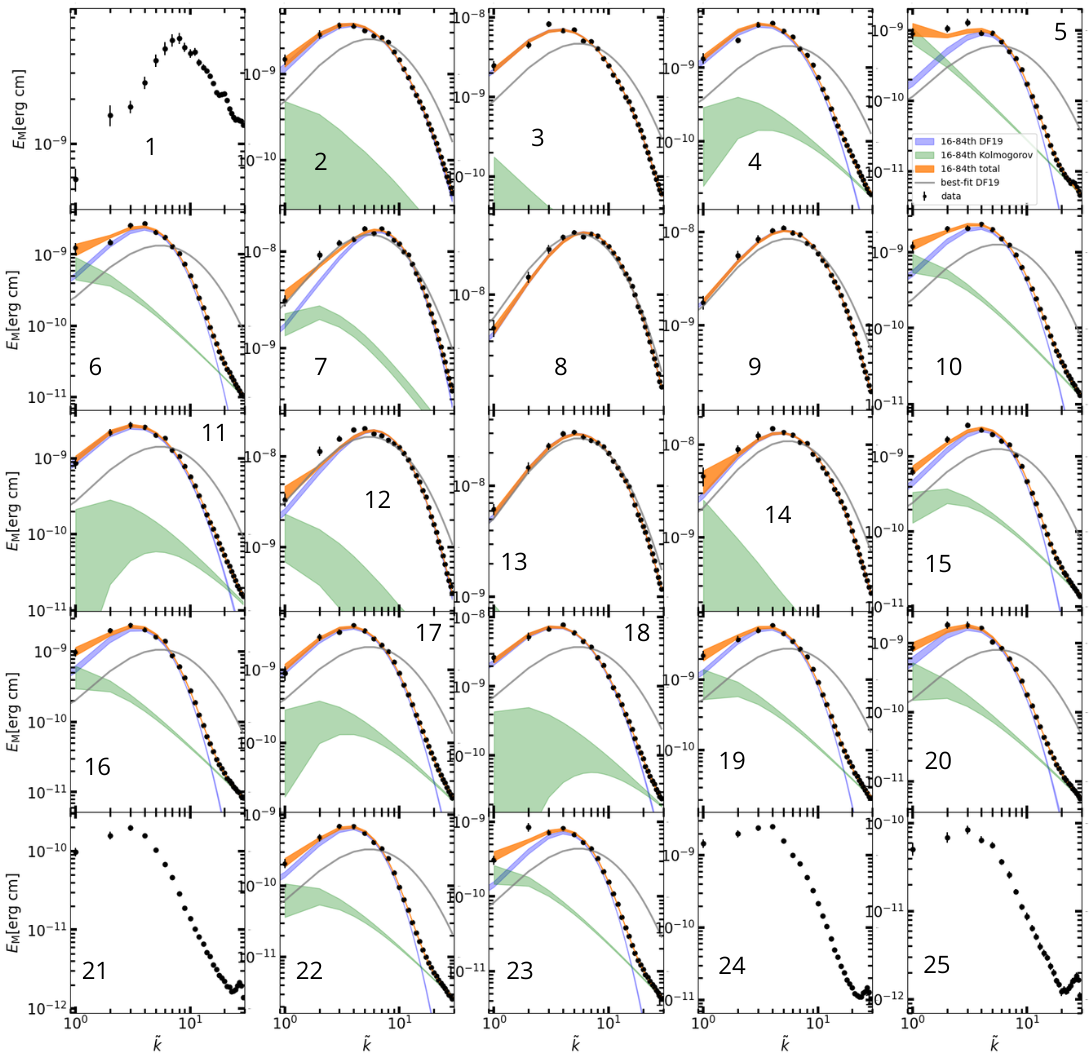}
    \caption{Same figure as \ref{fig:map_15}, but for the $\tilde k$ range 0 -- 30.
    In addition to the same DF19 model as in Fig. \ref{fig:map_15} (blue regions), we plot the total model (orange regions) and the Kolmogorov-like (green regions) model, respectively.
   }
    \label{fig:map_30}
\end{figure*}

\subsection{Results for the extended model ($D \neq 0$)}
\label{sec:posterior}

Fig. \ref{fig:map_30} shows the data of the 3D magnetic power spectrum in our subboxes over the $\tilde k$ range 1 -- 30.
The spectra of the peripheral subboxes contain numerical artifacts, such as spectral spikes.
For example, the data in the bottom right panel of Fig. \ref{fig:map_30} exhibit a prominent spectral spike approximately in the $\tilde k$ range 20 -- 30.
These spikes should occur just at values of $k$ larger than plausible values of $\kny$, whose rigorous determination is impractical due to the presence of nested AMR levels.
Given that this Section attempts to extend our analysis up to wavenumbers approaching the spectral spikes in peripheral subboxes, to avoid fitting these spikes that are not accounted for by Eq. (\ref{eq:model}), we preliminarily screen our subbox collection and visually reject 26 subboxes because of the presence of prominent spectral spikes at $\tilde k > 15$, in addition to the four subboxes rejected in Sect. \ref{sec:statistics}.
The rejected subboxes have only the data in Fig. \ref{fig:map_30} and in \myzenodo.

Leaving free the parameters $A$, $B$, $\lC$, $D$ and $F$ in Eq. (\ref{eq:model}) and assuming the same likelihood as in Sect.~\ref{sec:statistics}, we fit the model in Eq. (\ref{eq:model}) to the data of the magnetic power spectrum of the ICM in each subbox over the $\tilde k$ range 0 -- 30.
The parameter vector of the posterior, $\vec \theta$, is now composed by $\log (A /[10^{-9}\mathrm{erg \, cm^{-3}}] )$, $B$, $L /\lC$, $\log (E / A)$ and $\log ( L / \lF)$. 
The lower and upper bounds of the uniform priors on these parameters are reported in Tab. \ref{tab:fitting}.
Fig. \ref{fig:map_30} shows the comparison of our posterior model obtained in each subbox of our cluster central slice with the corresponding data of the 3D magnetic power spectrum (in \myzenodo \ we report the same figures as \ref{fig:map_30}, but for all the slices).
Because small numerical effects can still be present and affect a formal evaluation of the goodness of fit, the Bayesian p-value would not provide a fully physical assessment of model performance; instead, it would quantify the capability of our model to reproduce these residual numerical effects. 
Fig. \ref{fig:map_30} shows no clear deviation of our model in Eq. (\ref{eq:model}) from the 3D magnetic power spectrum over the $\tilde k$ range 1 -- 30 in the subboxes of our collection. 
Given that the statistical significance in reproducing the physics of peripheral subboxes is not very meaningful due to the presence of residual numerical effects in the simulation data, this indicates the "practical significance" of our model in describing the power spectrum of magnetic fluctuations down to a spatial scale of 30 kpc or up to $\tilde k = 30$.

%This Section first presents the results of the fitting of Eq. (\ref{eq:model}) to the data of the power spectrum of the magnetic field of the ICM in individual subboxes and, then, analyzes as a whole the parameter values inferred for all the subboxes.

%We now analyze this 75\% of the subboxes, which constitute our fiducial collection, given that repeating the same analysis (see below) but for this 51\% of subboxes yields results consistent with those of our fiducial collection.
For reference, in Fig. \ref{fig:map_30} we overplot separately the DF19 model and the Kolmogorov-like term\footnote{We obtain the DF19 model and the Kolmogorov-like term of Eq. (\ref{eq:model}) by fixing $D=0$ and $A=0$, respectively.}, according to the posterior of each subbox.
While the total (DF19 plus Kolmogorov-like) model overlaps with the DF19 model in most of the subboxes across a wide range of intermediate wavenumbers, in most of them the total model overlaps with the corresponding Kolmogorov-like term only at $\tilde k$ close to 30, implying that the latter is more relevant at large wavenumbers.
To measure individually for the posterior sampling of each subbox the deviation from the DF19 model, used in Sect. \ref{sec:statistics}, we consider the $k$ where the Kolmogorov-like term is more relevant, and compute the ratio between the median value of the latter at $\tilde k = 30$ and the corresponding median value of the total model.
About 90\% of the subboxes have this ratio higher than 0.1, implying that in most of our subboxes the data of the 3D power spectrum of the magnetic field are consistent with a significant Kolmogorov-like contribution at $\tilde k\approx 30$.
%This indicates that the decreasing trend of the Bayesian p-values moving towards the peripheral subboxes in Sect. \ref{sec:statistics} was due to the absence of a Kolmogorov-like component that in a few subboxes starts being relevant already at $kL \approx 15$.
The position of the cut-off, parameterized by $\lF$ in Eq.~(\ref{eq:model}), is often poorly constrained. 
Indeed, in many subboxes shown in Fig.~\ref{fig:map_30}, the 16th-percentile Kolmogorov-like spectrum exhibits a clear cut-off at intermediate wavenumbers, whereas the corresponding 84th-percentile spectrum shows no cut-off down to $\tilde k \approx 1$.
In many subboxes, the 16th-84th percentile interval spectra of both DF19 and Kolmogorov-like components are in a limited range of wavenumbers wider than that of the total model, implying that different combinations of these two terms yield, for these $k$, statistically indistinguishable total models.
This is a visual counterpart of the degeneracy in individual subboxes between the DF19 model parameters ($A$, $B$ and $\lC$) and those of the Kolmogorov-like term ($D$ and $\lF$).
However, the marginal posteriors of each parameter of individual subboxes are typically unimodal and quite symmetric around the median, except for the posterior of $\lF$ that is bimodal in a very few subboxes.
For each subbox, we evaluate the correlation between every pair of parameters using Spearman's rank correlation coefficient, $\rs$.
For each pair of parameters, considering the corresponding $\rs$ of all the individual subboxes, we build the distribution of all these $\rs$.
Spanning a relatively limited range of $\rs$ close to $\pm 1$, $A$-$B$, $A$-$\lC$, $A$-$D/A$ and $B$-$\lC$ are strong correlations present in almost all the posteriors of individual subboxes.

\begin{figure}
   \centering
   \includegraphics[width=0.499\textwidth]{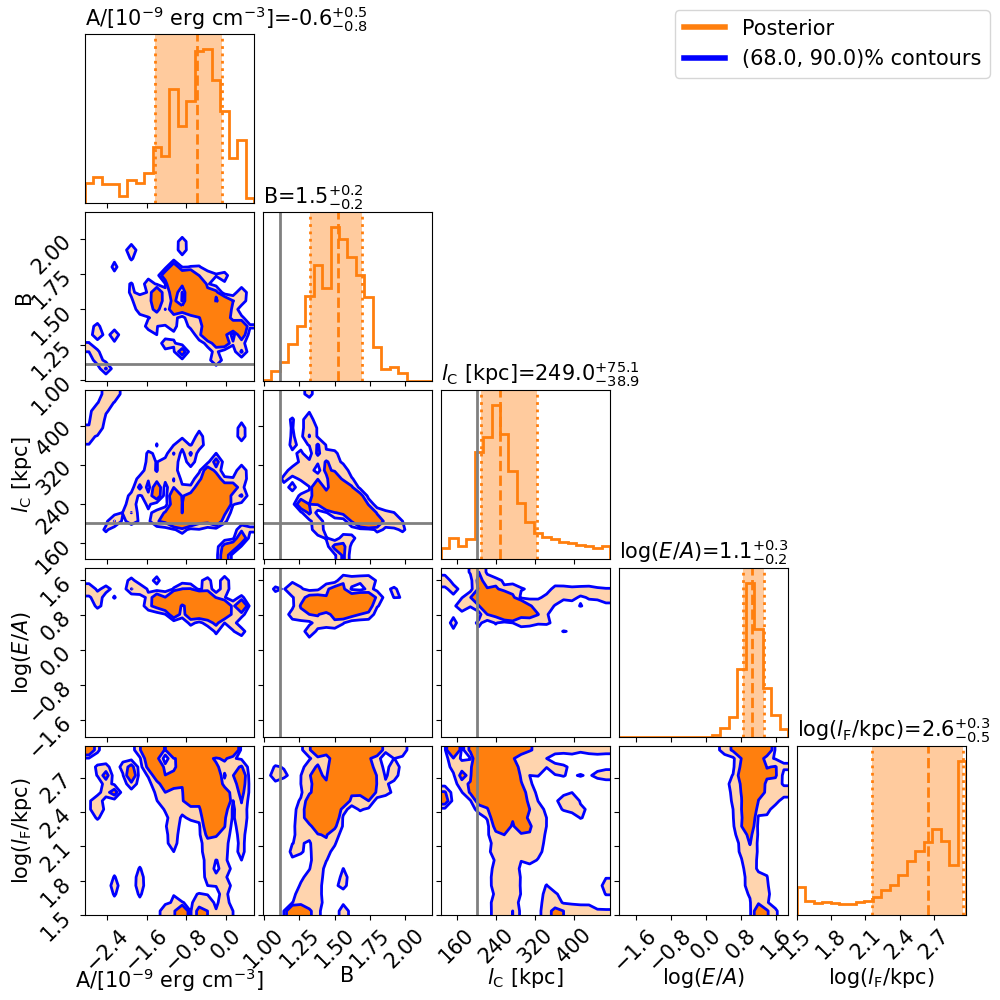}
    \caption{Marginal total posteriors (diagonal panels) and two-parameter joint posteriors (off-diagonal panels) of the Eq. (\ref{eq:model}) parameters ($A$, $B$, $\lC$, $D/A$, and $\lF$) for our subboxes defined for the $D \neq 0$ models in Sect. \ref{sec:posterior}.
    In each diagonal panel, the black curve and the orange histogram are the marginal prior and posterior, respectively;
    the light-orange vertical band is the 16th-84th percentile interval of the marginal posterior.
    In each off-diagonal panel, the dark-orange and light-orange regions, enclosed by the blue lines, define the 68\% and 90\% credible regions of the joint posteriors, respectively.
    The best-fit parameter values that define the shape of the 3D magnetic power spectrum in \citetalias{Dominguez19} are overplotted as gray lines.
   }
    \label{fig:corner}
\end{figure}

To study the distribution of our parameters across the entire cluster, we present the corner plot of the Eq.~(\ref{eq:model}) parameters in Fig.~\ref {fig:corner} for the total posterior sampling, comprising all posterior samples from the individual subboxes in our collection.
The marginal total posteriors of $A$, $B$, $\lC$, and $D/A$ are unimodal and quite symmetric around the median values, whereas the values of $\lF$ across the entire range of the uniform prior have a non-negligible posterior probability, with the most significant mode around 950 kpc.
As a summary of the posterior inference, we report the percentiles of the marginal posteriors for our parameters in Table \ref{tab:fitting}. %vector $\vec \theta$.
As shown in Fig. \ref{fig:corner}, the combination of the best-fit values of $B$ and $\lC$ of \citetalias{Dominguez19} lies outside the 90\% credible region (similar differences are also found for the total posterior of the $D=0$ fitting performed and analyzed in Sect. \ref{sec:statistics}).
When we compare the marginal posteriors of the $D=0$ and $D\neq 0$ models, both obtained for the subbox collection used for the latter model, we find that the median value of $B$ is higher by a factor of 1.14 in the $D=0$ models, while that of $\lC$ is lower by a factor of 0.86. 
In contrast, the median values of $A$ are similar between the two models. 
This indicates that both the model complexity and the $\tilde{k}$ range used in the fit affect the inferred values of $B$ and $\lC$.
We define for a given parameter of $\vec \theta$ its 1$\sigma$ relative uncertainty as the ratio between its 16th-84th percentile interval and its median value.
The ratio, $\mathcal R$, between the median of the 1$\sigma$ relative uncertainty of each parameter among all the subboxes of our collection and the 1$\sigma$ relative uncertainty from the marginal total posterior is reported in Table \ref{tab:fitting} (values similar to those listed there are obtained in the $D=0$ fitting of Sect. \ref{sec:statistics}).
All our parameters, except for $\lF$, have values of $\mathcal R$ in the range 0.05 -- 0.2, suggesting that the subbox-to-subbox variations in all these parameters are much more significant than the typical uncertainty in the parameter estimation in individual subboxes and dominate the scatter reported in the marginal posterior of Fig. \ref{fig:corner}.
This implies that, in the modeling of the observed RM maps, fixing the parameters of a model for the 3D magnetic power spectrum (whether that of \citetalias{Dominguez19} or any other) and adopting throughout the cluster volume the corresponding shape just cannot account for the entire observed diversity of spectra in the simulated cluster volume. 
This, in turn, may lead to inaccurate constraints on the profile of the magnetic field strength (see Sect. \ref{sec:intro}).
This result further motivates our investigation of the physical drivers of this subbox-to-subbox diversity in the following Section.

%Now we compare the results of our fitting to those of \cite{Dominguez19} for E18B, except for the normalization of the Kazantsev model $A$, because it is associated with the energy of the magnetic fluctuations within the subbox and thus depends on the subbox side length.
%The \cite{Dominguez19} best-fit parameter corresponding to $\lC$ of Eq. (\ref{eq:model}), 4.72, falls within the 16th-84th percentile interval, whereas B of \cite{Dominguez19}, 1.11, lies below the 16th percentile, but within the 5th-95th percentile interval.
%Eq. (4) of \cite{Dominguez19}, with its free parameters fixed to their best-fit values for E18B, is thus recovered in our analysis as a particular case of our more general model.

%The similarity of parameter pair correlations in the total posterior, plotted in the off-diagonal panels of Fig. \ref{fig:corner}, to those typically present in individual subboxes, reported in ..., indicates that the correlations in the total posterior reflect those in the posterior of individual subboxes.

\section{Dependence of the magnetic power spectrum on local properties}
\label{sec:correlation}
This Section attempts to identify the likely physical drivers of the measured subbox-to-subbox diversity in the 3D power spectra of the magnetic field of the ICM (see Sect. \ref{sec:posterior}).
Our simulated cluster can be considered a prototype of a post-merging cluster that underwent its last major merger at $z \approx 0.5$.
We work under the simplifying assumption that the 3D power spectrum of the magnetic field depends on local ICM properties, even though neighboring fluid elements at $z=0$ can originate from widely separated regions at high $z$ and therefore likely underwent different evolutionary histories.
Specifically, using Spearman's correlation coefficient ($\rs$), we examine the statistical significance of a monotonic relation between the inferred values of each term in Eq. (\ref{eq:model}) and physical parameters of the same snapshot and subboxes as for the corresponding data of the 3D magnetic power spectrum.
Given that we resolve the spatial variations of the Eq. (\ref{eq:model}) parameters on scales of the subbox size length, in Sect. \ref{sec:appendix}, we evaluate a generic gas property $Q$ in each sub-box as the average, $\bar Q$, within each subbox defined in Sect. \ref{sec:fourier}.

\begin{figure*}
   \centering
   \includegraphics[width=0.49\textwidth]{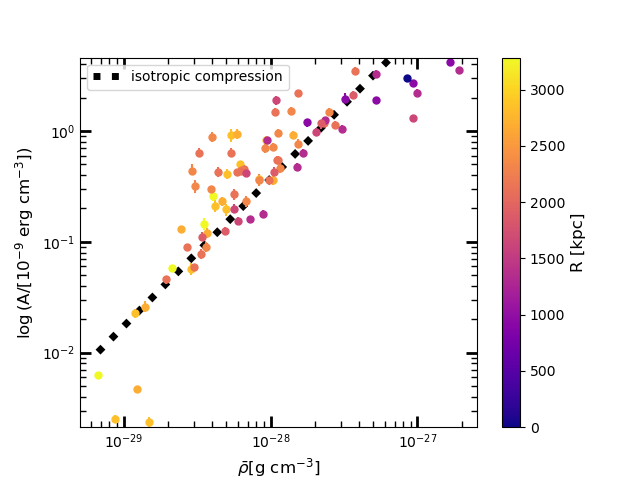}
   \includegraphics[width=0.49\textwidth]{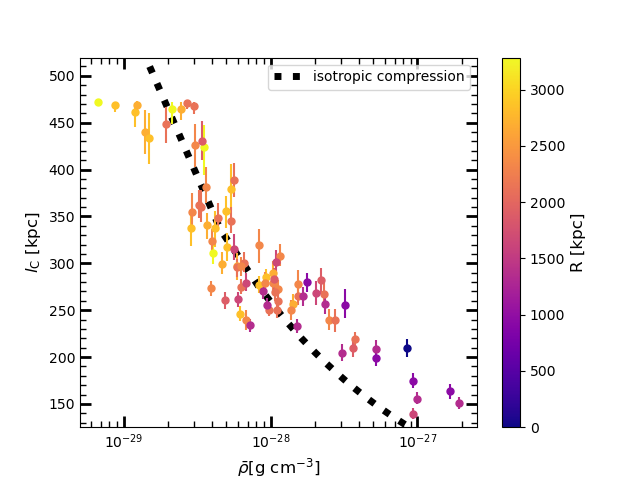}
    \caption{Scatter plots of our model parameters $A$ (left panel) and $\lC$ (right panel; see Eq. (\ref{eq:model})) with the average density within a subbox, $\bar \rho$ (see Appendix \ref{sec:dynamics}, for details). 
    The data points and the error bars are the median values and the 16th-84th percentile interval in the $D=0$ models, selected based on $t=0.01$ (see Sect. \ref{sec:statistics}). 
    The black dotted lines are the expected scalings in the model of adiabatic, isotropic compression (see text).
    Each data point is color-coded by the distance $R$ of its subbox from the central one.}
    \label{fig.corr_A_lC}
\end{figure*}

In Appendix \ref{sec:dynamics}, we describe the quantities $\bar Q$ and their importance for the fluctuation dynamo or cluster turbulence.
We, then, evaluate $\rs$ among these $\bar Q$ that we report in Fig. \ref{fig:cross}, to quantify how much the quantities $\bar Q$ depend on each other. 
As expected, many $\bar Q$ significantly correlate with the density, as discussed in Appendix \ref{sec:dynamics}.
In Appendix \ref{sec:degree}, we describe the statistical method to evaluate the significance of the correlation of each pair composed of one of the Eq. (\ref{eq:model}) parameters and one of these $\bar Q$.
Figs. \ref{fig:corr_dynamo} and \ref{fig:corr_extended} show, for the $D=0$ and $D\neq 0$ models, respectively, the correlations between our parameters and these $\bar Q$, quantified by $\rs$ as detailed in Appendix \ref{sec:degree}. 
All the parameters except for $\lF$ ($A$, $B$, $\lC$ and $D/A$) in one of the two models have $|\rs| > 0.3$ with one or more $\bar Q$. 
For each of the parameters $A$, $B$, $\lC$, and $D/A$, we apply the procedure described in Appendix \ref{sec:degree} to disentangle direct from indirect correlations and to identify the main drivers of the subbox-to-subbox variations.
The main results of this procedure are discussed in detail here, whereas the values of $\rs$ for all the quantities under consideration are reported in Figs. \ref{fig:corr_dynamo} and \ref{fig:corr_extended} for the $D=0$ and $D\neq 0$ models, respectively.

%Fig. \ref{fig.corr_A_lC} shows that the normalizations of the Kazantsev and Kolmogorov-like components of Eq. (\ref{eq:model}), $A$ and $D$, increase with increasing ICM density $\rho$, with $\rs \approx$ 0.88 and 0.8, respectively.

Given that the spectra in the $\tilde k$ range 1 -- 15 are more robust against numerical effects, we use as a baseline the correlations of the parameters $A$, $B$, and $\lC$ for the $D=0$ models and, then, report the values of $\rs$ for the $D\neq 0$ models.
Fig. \ref{fig.corr_A_lC} shows the scatter plots of the two parameters of the DF19 model, $A$ and $\lC$ (see Sect.~\ref{sec:model}), obtained from the dynamo-like ($D=0$) models presented in Sect.~\ref{sec:statistics}, as a function of the ICM density for the $t=0.01$ subbox collection (see Sect.~\ref{sec:statistics}). 
We find strong monotonic correlations, with $\rs\approx 0.87$ for $A$ and $\rs\approx-0.85$ for $\lC$ (similar values are obtained for $t=0.05$). 
The extended ($D\neq0$) models, built in Sect.~\ref{sec:statistics}, yield consistent trends, with $\rs\approx0.92$ for $A$ and $\rs\approx-0.54$ for $\lC$ (see Fig. 31 in \myzenodo).
Since the magnetic energy is given by the 1D integral $\int E_{\rm M}(k)dk$, the parameter $A$ sets the magnetic energy density of the asymptotic regime of the DF19 model (see Sect. \ref{sec:model}, for the definition) at low wavenumbers within each subbox, whereas $\lC$ characterizes the physical scale of the fluctuations containing the largest fraction of the magnetic energy (see Eq. (\ref{eq:model})).
These correlations suggest that, as the ICM density increases, the magnetic energy increases and becomes increasingly concentrated on smaller scales.

We can interpret these correlations in terms of the simple, isotropic, adiabatic compression of a magnetized plasma in the stratified ICM. 
When a bundle of magnetic field lines in a uniform plasma is compressed, by mass and magnetic flux conservation, the magnetic energy density increases with increasing ICM density, $\rho$, as $B^2 \propto \rho^{4/3}$ and the typical correlation scale of the magnetic fluctuations shrinks approximately as $\rho^{-1/3}$ \citep[see Sect. 3.1 of][]{Donnert18}.
These model-predicted scalings are compared with the simulation results in Fig.~\ref{fig.corr_A_lC}. 
The agreement between the measured trends and these theoretical scalings is good. 
However, the latter are obtained for an idealized model in which, at the start of compression, the magnetic and density fields are uniformly distributed throughout the cluster volume. 
In reality, they likely began with a patchy distribution of differentially amplified regions after the cluster had fully formed.

These findings are also consistent with those reported by \citetalias{Dominguez19}, who interpreted the evolution of the best-fit parameters of the magnetic spectra in the DF19 model according based on compression\footnote{See also the recent results of \citet{Abramson25, Schober26} on the role of compression in triggering a forward cascade of primordial magnetic energy during the formation of the large-scale structure.} as described above. 
In this scenario, merger-driven compression transfers magnetic energy from larger to smaller scales, leading to a reduction of the magnetic correlation length.
Specifically, the decrease in $\lC$ coincident with the merger events \citepalias[see Fig.~18 of][]{Dominguez19} supports this interpretation.

%The connection between the magnetic energy concentration on small scales and the deepening of the gravitational potential well has also been investigated in the broader context of magnetic field amplification during the formation of large-scale structure. 
%In particular, \citet{Abramson25} developed a theoretical model describing the forward cascade of primordial magnetic energy during structure formation, reproducing, at least in part, the transfer of this primordial magnetic energy toward smaller spatial scales observed in cosmological simulations of structure formation \citep[e.g.][]{Schober26}.

\begin{figure}
   \centering
   \includegraphics[width=0.49\textwidth]{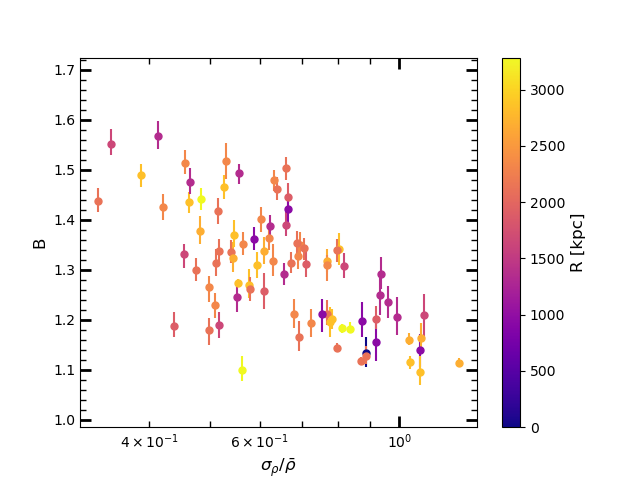}
    \caption{Scatter plots of our model parameter $B$ (see Eq. (\ref{eq:model})) in the $D=0$ models as a function of our proxy for ICM clumpiness, $\sigma_\rho / \bar \rho$ defined in Appendix \ref{sec:dynamics}. The data points, their error bars and the color-coding are as in Fig. \ref{fig.corr_A_lC}.}
    \label{fig.corr_B}
\end{figure}

\begin{figure*}
   \centering
   \includegraphics[width=0.49\textwidth]{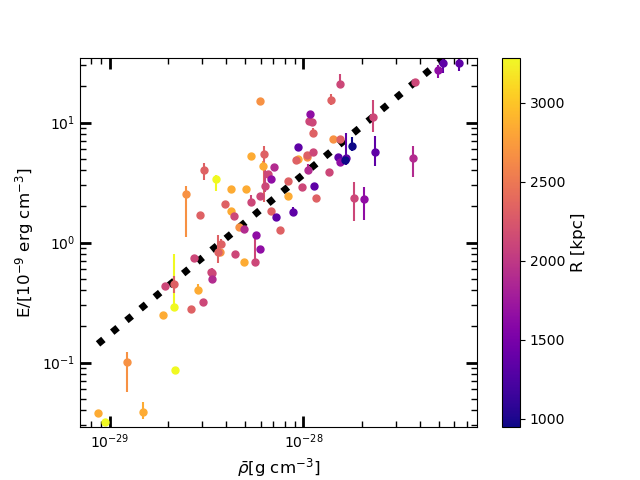}
   \includegraphics[width=0.49\textwidth]{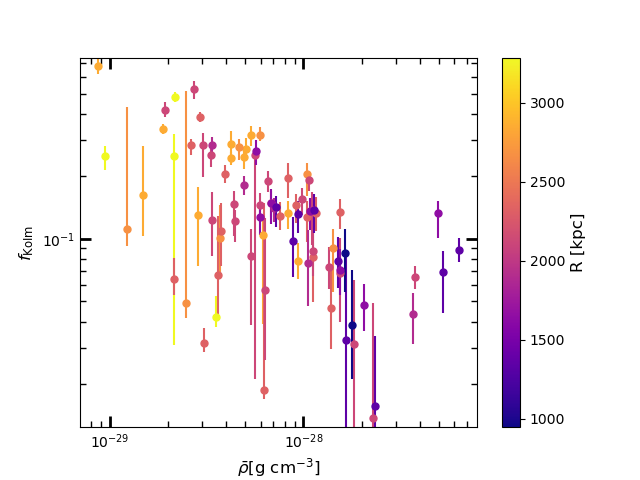}
    \caption{Scatter plots of our model parameter $D$ (see Eq. (\ref{sec:posterior}); left panel) and of the energy ratio $\fkol$ (see text; right panel) as a function of the ICM density, $ \bar \rho$, for the $D\neq 0$ models. The data points, their error bars, the scaling, and the color-coding are defined as in Fig. \ref{fig.corr_A_lC}.}
    \label{fig.corr_E/A}
\end{figure*}

Fig. \ref{fig.corr_B} shows the scatter plot between our parameter $B$ and our proxy for the degree of gas clumping in the ICM, defined in App.~\ref{sec:appendix} as the ratio between the standard deviation ($\sigma_\rho$) and the mean ($\bar \rho$) of the density distribution within a given subbox, for the dynamo-like ($D=0$) models selected based on $t=0.01$. 
We find a significant anticorrelation of $\rs\approx-0.57$, indicating that $B$ decreases as the ICM distribution becomes increasingly clumpy within a subbox (a similar $\rs$ is obtained for the $t=0.05$ subbox collection). 
For the extended ($D\neq0$) models, however, the correlation weakens substantially down to $\rs \approx -0.34$ (see Fig. 32 in \myzenodo), similarly to the $\lC$-$\rho$ correlation.
%The increase in the number of free parameters in the extended models introduces fitting complexity that appears unnecessary in the wavenumber range sensitive to the parameter $B$ (approximately $\tilde k < 15$), leading to additional noise in the estimation of $B$.
%This could lead to a reduction in the correlation significance.
As shown in the top panel of App.~\ref{fig.varying}, increasing $B$ produces a narrower peak in the DF19 model \citepalias[see also Fig.~1 of][]{Dominguez19}. 
Consequently, fluctuations that contain the largest fraction of magnetic energy are distributed over a wider range of wavenumbers in the presence of prominent density substructures.
\citetalias{Dominguez19} observed an ambiguous trend of $B$ with the assembly history of the cluster, leaving room for different interpretations, including a scenario in which a lower $B$ is associated with the more active phases of the assembly history of the clusters characterized by infalling clumps.
If confirmed, this interpretation would be consistent with what we provide for the above correlation.

The left panel of Fig. \ref{fig.corr_E/A} shows the scatter plots of our parameter $D$ as a function of the ICM density.
We find a correlation with $\rs \approx 0.72$, implying that the normalization of the asymptotic behavior of the Kolmogorov-like term of our 3D power spectrum model at large wavenumbers increases with increasing ICM density.
Given that $D$ determines the energy density of magnetic structures not in a fully dynamo regime, we compare this relation with the scaling predicted by our adiabatic, isotropic model for the magnetic energy density described above and find good agreement.
In combination with the agreement of the dependence of $A$ and $\lC$ on ICM density, this indicates that the adiabatic compression of the ICM significantly shapes all the magnetic structures throughout the cluster volume.

In the Appendix \ref{sec:degree}, we also find evidence for other possible but weaker correlations between fitting parameters and local ICM properties. 
Fig. \ref{fig:corr_extended} presents a correlation between the fitting parameter $D/A$ and the ICM density $\rho$, with $\rs \approx 0.57$.
Rather than using the ratio of the normalizations of the asymptotic DF19 and Kolmogorov-like components, we adopt the following quantity, which is more directly relevant to interpreting the results of our correlation analysis.
For each parameter vector $\vec \theta$ in our posterior sampling, we evaluate the ratio, $\fkol$, between the total magnetic energy obtained after integrating the power spectra 
of the Kolmogorov-like model (obtained with $A=0$ in Eq. (\ref{eq:model})) and of the full total model given by Eq. (\ref{eq:model}), respectively.
The right panel of Fig. \ref{fig.corr_E/A} shows the scatter plots of $\fkol$ as a function of the ICM density.
We find a correlation of $\rs \approx - 0.67$, implying that the energy stored in the magnetic fluctuations that are less amplified by the fluctuation dynamo increases as we move towards the low-density regions of our cluster.

Figure \ref{fig:corr_extended} presents tentative evidence for a correlation between $D/A$ and the Mach number of the velocity fluctuations, $\mathcal{M}_\sigma$, with $\rs \approx 0.47$, although this trend is not readily evident in the scatter plot of Fig. 33 in \myzenodo. 
Using $\fkol$ instead of $D/A$ significantly weakens this correlation.
Such a correlation is qualitatively consistent with the results of turbulence-in-a-box simulations, which predict that the fluctuation dynamo becomes progressively less efficient with increasing $\mathcal{M}_\sigma$ \citep{Federrath11, Schober15}. 
In the turbulence-in-a-box simulations, $\mathcal{M}_\sigma$ is approximately constant throughout the evolution, and a larger $\mathcal{M}_\sigma$ primarily delays the growth of the magnetic field and the development of its folded structure during the kinetic and linear phases of the fluctuation dynamo \citep{Federrath11, Schober15, Sur24}. 
Comparing our correlation between $D/A$ and $\mathcal M_\sigma$ with the results of these turbulence-in-a-box simulations would therefore require following in our cosmological simulation the evolution of $\mathcal{M}_\sigma$ of the fluid elements within each subbox over their entire history. 
The relatively weak correlation we find may simply reflect that our analysis is based solely on the $z=0$ properties and does not account for the cumulative turbulent history of these fluid elements over cosmic time.
The lack of significant correlations with our proxies for departures from volume-filling, isotropic MHD turbulence, typical of fully controlled numerical experiments, may indicate either that the 3D magnetic power spectrum is primarily determined by the integrated cluster assembly history, as discussed above, or that the fluctuation dynamo is only weakly sensitive to deviations from isotropic, volume-filling MHD turbulence occurring on scales larger than where the fluctuation dynamo should act.

In summary, we find that the correlations $A$-$\rho$, $B$-$\sigma_\rho/\rho$, $\lC$-$\rho$ and $\fkol$-$\rho$ account for a substantial fraction of the subbox-to-subbox variations. 
This suggests, therefore, that the variations are driven by the local environment and by the local dynamics of the ICM. 
Such dependencies on the ICM conditions are currently not included in the RM modeling performed in observations, which, even in the best case, resort to the sophisticated best-fit DF19 model and assume it throughout the entire cluster extent. 
Therefore, we suggest that the findings of this Section can be used to mitigate this limitation. 
As an example of a possible implementation in the observational analyses, we propose the following procedure. 
(i) The observer estimates the average density of the intervening ICM along the LOS in the foreground of a given polarized source used for the cluster RM mapping. 
(ii) Using the correlations shown in Figs. \ref{fig.corr_A_lC} - \ref{fig.corr_E/A}, the observer associates this density with the corresponding values of the $D\neq0$ model parameters. 
(iii) For the RM map of that source, the observer then adopts the 3D magnetic power spectrum described by our proposed functional form, with the parameters fixed to the values inferred in the previous step. 
(iv) The same procedure is repeated independently for each RM map. 
Compared with the common assumption of using the same parameter values across the entire cluster extent, this approach approximately accounts for local variations and is therefore expected to provide a more realistic description of the magnetic field, according to our analysis. 
This would be the first attempt to generate a physically motivated 3D model of magnetic fields in the ICM, without assuming efficient small-scale dynamo amplification at all radii. 
Of course, after establishing the formalism in this paper, more work is needed to generalize and validate the approach against a larger sample of simulated galaxy clusters, and to assess likely additional dependencies on the cluster dynamical state.

\section{Conclusions}
\label{sec:conclusions}

We have presented a new study on how the magnetic power spectrum is expected to vary across the ICM as a function of the local environment and dynamics.
We analyzed a high-resolution snapshot of a simulated cluster at $z=0$, produced with a zoom-in AMR cosmological simulation with the ENZO code \citep{Vazza18_magne}. 
Such a theoretical analysis has a strong connection with observations, since the steady improvement in the sensitivity and coverage of radio surveys makes it possible to use the Faraday rotation effect on polarized radio sources as a tool to map magnetic fields out to large cluster radii \citep[e.g.][]{Enblin03,Bonafede10, Bonafede13, Bonafede15, Stuardi21, DeRubeis24, Alonso26, Loi26}. 
 However, the reconstruction of the 3D distribution of magnetic fields from Faraday rotation data in the previous work is often degenerate with respect to the combination of the thermal electron density, the strength, and the scale distributions (i.e., its power spectrum) of magnetic fields. %\citep[e.g.][]{Bonafede13, Stuardi21,Alonso26}. 
 Therefore, a physically motivated model to predict how the magnetic power spectrum may change with the local properties of the ICM, as in this new work, is highly relevant. 

Our simulated cluster has been the object of previous studies presenting evidence for small-scale dynamo throughout an extended region of the cluster and magnetic properties consistent with RM observations of the Coma cluster \citep{Vazza18_magne, Dominguez19}.
In this work, after partitioning the cluster volume into 125 identical subboxes of side length $\approx$ 950 kpc, we derived their magnetic and kinetic power spectra (see Sect. \ref{sec:simulation}) and attempted to describe them with a flexible fitting formula.
Our starting point is the fitting formula for the magnetic field spectrum by \citetalias{Dominguez19}, originally proposed and validated in the central region of simulated clusters.
We confirm that this formula provides an excellent fit to the outcome of small-scale dynamo amplification in the ICM, as we found in the central subboxes or in magnetic fluctuations on spatial scales larger than 60 kpc (see Sect. \ref{sec:statistics}). However, 
we expanded this functional form by adding  %upon this idea in two ways: a) we monitored how the best-fit parameters in the DF19 model change across the simulated ICM and b) we added 
a Kolmogorov-like term to account for the turbulent stirring of magnetic fluctuations in regions where the dynamo process is inefficient, either for physical or numerical reasons (see Sect. \ref{eq:model}).
We then fit our extended model for the magnetic power spectrum to the corresponding data in each subbox using an MCMC algorithm (see Sect.~\ref{sec:results}).
We eventually investigate how the inferred parameters vary across the simulated ICM, under the assumption that the magnetic power spectrum in each subbox is determined solely by the local physical conditions within that subbox.
Specifically, 
%To identify the dependence of the parameters of our extended model under the simplistic assumption that the magnetic power spectrum depends on the local ICM conditions, 
we examine the scatter plots of the inferred parameter values with the corresponding ICM properties (see Sect. \ref{sec:correlation}).

Our main findings are the following.
\begin{itemize}
    \item The best-fit magnetic power spectrum of \citetalias{Dominguez19} is appropriate to describe the corresponding data limited to the central ($\approx$ 1 Mpc$^3$) subbox of our cluster (see Sect. \ref{sec:statistics}). 
    \item Our extended model for the magnetic power spectrum well reproduces the corresponding data in most of our individual subboxes, with inferred parameter values that vary significantly from subbox to subbox (see Sect. \ref{sec:posterior}).
    \item The correlations of our model parameters with density and clumpiness of the ICM can account for a large fraction of this subbox-to-subbox diversity in the magnetic power spectrum (see Sect. \ref{sec:correlation}).
\end{itemize}

However, the analysis presented in this paper has some limitations and could be expanded in several ways in future work. 
First, we focused only on a single snapshot of a single cluster, though representative of clusters where the RM observations are most likely (see Sect. \ref{sec:intro}), and did not address whether and how much our findings change with varying spatial resolution of the simulation and different MHD solvers. This requires analysis of a larger body of simulations, possibly including the local contribution of AGN to RM in clusters \citep[e.g.][]{2011ApJ...739...77X,lehle2026A&A...705A..41L}, which we defer to future work. 

Second, the implications of these findings for modeling RM maps derived from radio observations of galaxy clusters (see Sect. \ref{sec:intro} for details) can be better explored through dedicated work employing mock radio observations.
Recent works, such as \citet{Bonafede13,Stuardi21,DeRubeis24, Alonso26, Loi26, Bon26}, fixed the shape of the power spectrum to the best-fit model of \citetalias{Dominguez19} in the outskirts of galaxy clusters or beyond.
%Our analysis showed this assumption is progressively less appropriate moving towards the outskirts of our simulated cluster (see Sect. \ref{sec:statistics}). %, where the best-fit magnetic power spectrum of \citetalias{Dominguez19} provides an appropriate description only for projected distances from the cluster center approximately smaller than 1 Mpc (see Sect. \ref{sec:statistics}). 
%For example, adopting the best-fit spectrum of \citetalias{Dominguez19} for the modeling of the mock RM maps taken from the outskirts of our simulated cluster is likely to underestimate the contribution of the magnetic fluctuations on small scales (modeled with a Kolmogorov-like term in Sect. \ref{sec:model}).
Our analysis indicates that the small-scale magnetic fluctuations in individual subboxes are better captured by an extended model that adds a Kolmogorov-like term to the DF19 model (Sect. \ref{sec:model}).
This addition becomes progressively more relevant towards the outskirts of our simulated cluster (see Sect. \ref{sec:posterior}). 
We suggest that future RM studies should consider this additional component when modeling the magnetic power spectrum.
Though a full assessment of the impact of these results on the modeling of RM maps is beyond the scope of this work, we note that our extended model may enhance the predicted RM scatter in the outskirts. 
This potentially leads to flat RM scatter profiles similar to those reported by \citet{Alonso26}, who studied the intercluster bridge and region (0.3 -- 1.8) $r_{500}$ of the Shapley Supercluster Core.
%Though an assessment of the impact of these inappropriate assumptions on the modeling of RM maps is beyond the scope of this work, it is worth mentioning that assumptions on the shape of the magnetic power spectrum different from the best-fit of \citetalias{Dominguez19} may alleviate the tension with the cosmological simulations claimed in some papers, such as \citet{Alonso26}.
%When the RM maps discreetly sample the entire cluster extent in the plane of the sky and the analyses want to assume one power spectrum across the entire cluster extent, it is more appropriate to extract the parameter from the posterior distribution of our fiducial subbox collection, shown in Fig. \ref{fig:corner}.
According to our findings, the main caveat for the observational modeling of RM maps, aimed at accurately constraining the cluster magnetic field strength profile, is the subbox-to-subbox diversity of the magnetic power spectrum. 
A statistically robust treatment of this diversity requires propagating the scatter in the model parameters into the uncertainties on the inferred magnetic field strength profile.

The likely physical connection of our model parameters with observationally accessible ICM quantities, which can be obtained from X-ray observations with current (e.g., XMM-Newton\footnote{\hyperref[https://www.cosmos.esa.int/web/xmm-newton]{https://www.cosmos.esa.int/web/xmm-newton}}, Chandra\footnote{\hyperref[https://chandra.harvard.edu/]{https://chandra.harvard.edu/}} and XRISM\footnote{\hyperref[https://www.xrism.jaxa.jp/en/]{https://www.xrism.jaxa.jp/en/}}) or future facilities (e.g., HUBS\footnote{\hyperref[https://hubs-mission.cn/en/index.html]{https://hubs-mission.cn/en/index.html}}, NewAthena\footnote{\hyperref[https://www.the-athena-x-ray-observatory.eu/en]{https://www.the-athena-x-ray-observatory.eu/en}}), offers the possibility of constraining part of the scatter in our model parameters from independent observations, thereby mitigating this caveat.
For example, \citep{Bot26} recently used a similar numerical simulation to explore a possible physical connection between the projected distribution of magnetic field vectors derived from fluctuations in synchrotron emission and the structure of the ICM velocity field in the same regions.  

In conclusion, our findings represent an important step towards better exploitation of the information encoded in the RM maps obtained with current and forthcoming radio interferometers, and we underscore the benefit of synergy among radio observations, X-ray observations, and numerical modeling of galaxy clusters, particularly in the forthcoming SKA era.

%the time evolution properties are inaccessible to observations

%\begin{figure*}
   %\centering
   %\includegraphics[width=0.33\textwidth]{Figures/only_rotating_ICM_joint_X_SZ_datasets_Abell85_Norm.png}  
   % \caption{}
    %\label{fig.A85}
%\end{figure*}
\begin{acknowledgements}
TB acknowledges funding from the European Union NextGenerationEU.
FV acknowledges funding under the European Union’s Horizon Europe program through the ERC Synergy Grant COSMOMAG (Project Id. 101224803). FV acknowledges the CINECA award  ``IscrB{\_}CREW"  under the ISCRA initiative, for the availability of high-performance computing resources and support, and the usage of online storage tools kindly provided by the INAF Astronomical Archive (IA2) initiative (http://www.ia2.inaf.it).
SE acknowledges the financial contribution from {\it Theory Grant / Bando INAF per la Ricerca Fondamentale 2024} on ``Constraining the non-thermal pressure in galaxy clusters with high-resolution X-ray spectroscopy'' (1.05.24.05.10).
\end{acknowledgements}

\section*{Data availability}
Appendices C and D are available on the Zenodo repository at the link
\url{https://doi.org/10.5281/zenodo.21870521}. 

\bibliography{refs}{}
\bibliographystyle{aa}

\begin{appendix}
\section{Model flexibility}
\label{sec:flexibility}
\begin{figure}
   \centering
   \includegraphics[width=0.4\textwidth]{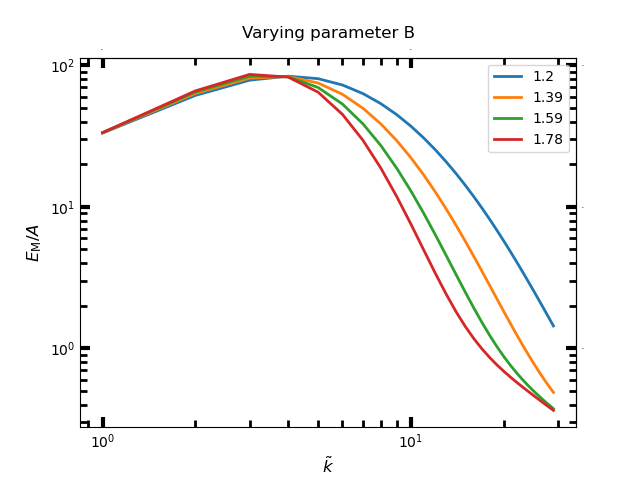}
   \includegraphics[width=0.4\textwidth]{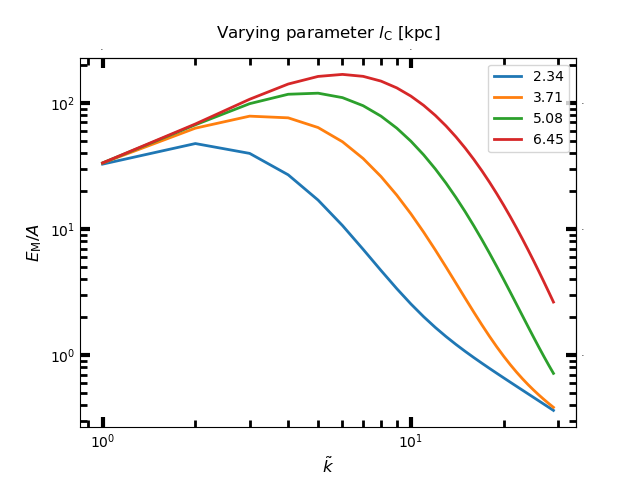}
   \includegraphics[width=0.4\textwidth]{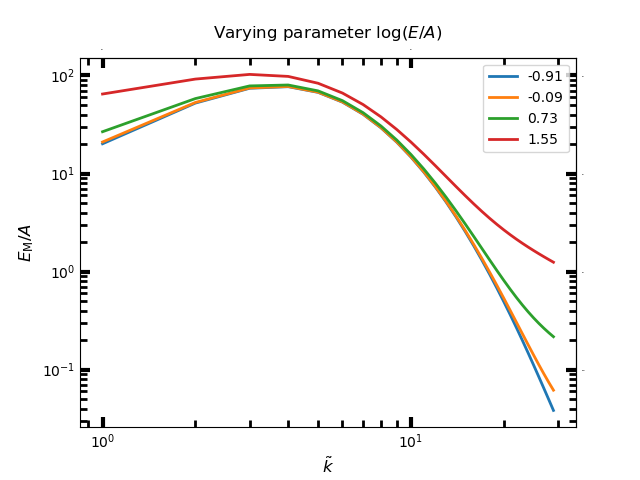}
   \includegraphics[width=0.4\textwidth]{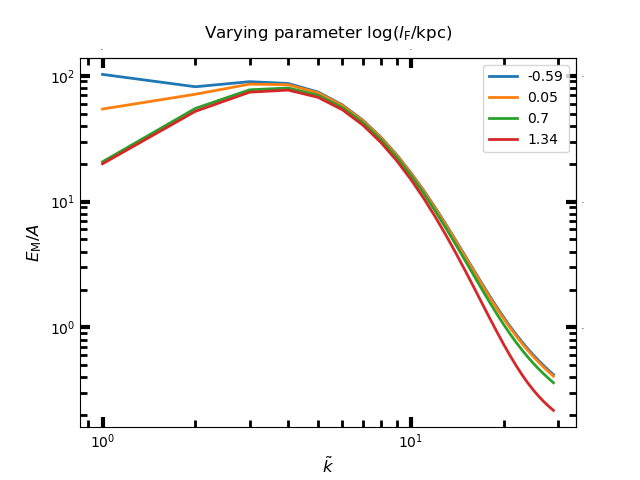}
    \caption{Effect of varying the values of the parameters in Eq.~(\ref{eq:model}): $B$ (top), $L / \lC$ (upper middle), $\log (E/A)$ (lower middle), and $\log (L / \lF)$ (bottom). 
    In each panel, only the displayed parameter is varied linearly across its 5th–95th percentile interval of the $D\neq 0$ models, while all others are fixed to their median values reported in Tab. \ref{tab:fitting}. 
    The values of the varying parameter in each plot are reported in the legend. 
    $A=1$ in all the plots.}
    \label{fig.varying}
\end{figure}

Figure \ref{fig.varying} shows a collection of power spectra of the magnetic field obtained from the model in Eq. (\ref{eq:model}) for a representative combination of values of the parameters $B$, $\lC$, $D$, $D$ and $\lF$ lying within their 5th-95th percentile marginal posterior interval inferred in Sect. \ref{sec:results} and reported in Table \ref{tab:fitting}. 
This gives a visual impression of the exact role played by our fitting parameters in modifying the magnetic power spectrum.

\section{Correlation analysis}
\label{sec:appendix}
This appendix presents the statistical analysis of the scatter plots of our inferred parameters with the ICM properties, whose main results are described in Sect. \ref{sec:correlation}.
Sect. \ref{sec:dynamics} and \ref{sec:degree} describe the considered dynamical properties and the method to quantify the significance of possible correlations, respectively.

\subsection{Dynamical properties of the ICM turbulence}
\label{sec:dynamics}
We here report the quantities evaluated within the individual subboxes, $\bar Q$, used to explore the dependencies between the parameters of the magnetic power spectrum model in Eq. (\ref{eq:model}) and the local ICM conditions.
Throughout this work, barred quantities indicate average quantities within individual subboxes.
The $\bar Q$ we consider for this study are the following.

\begin{itemize}
    \item The plasma parameter $\betatherm = 8 \pi \bar \rho \kboltz \bar T / (\mu \proton \bar{B^2})$ measures the dynamical importance of the magnetic pressure with respect to the thermal pressure of the ICM. 
    Here $\rho$, $T$ and $B$ are the density, temperature and magnetic field strength, respectively, and $\kboltz$, $\mu = 0.6$ and $\proton$ are the Boltzmann constant, the molecular weight and proton mass, respectively.
    \item The radial distance of the subbox from the cluster center (see Sect. \ref{sec:simulation}), $R$, traces the position of the subbox center, whereas the average mass density within the subbox, $\bar \rho$, the subbox where the ICM has been compressed more significantly. 
    \item We define the Mach number of velocity fluctuations within a given subbox as $\mathcal M_\sigma = \sqrt{\sigma_x^2 + \sigma_y^2 + \sigma_z^2} / \sound$, where $\sigma_h$, with $h = \{x, y, z\}$, is the Cartesian diagonal $h$-component of the turbulent velocity dispersion tensor and $\sound = \left( 5 \kboltz \bar T / (3\mu \proton)\right)^{1/2}$.
    As shown in turbulence-in-a-box simulations by \citet{Federrath11}, $\mathcal M_\sigma$ significantly contributes to determining the efficiency of the fluctuation dynamo, with subsonic turbulence ($\mathcal{M}_{\sigma} \leq 1 $) being more efficient in the amplification compared to supersonic turbulence ($\mathcal{M}_{\sigma} > 1 $).
    \item From the Helmoltz decomposition \citep[e.g.,][]{valles24} it follows that the total velocity field of the ICM, $\vec \utot$, can be written as  $\vec \utot = \vec \ucomp + \vec \usol$, where $\vec \usol$ is the divergence-free ($\nabla \cdot \vec \usol= 0$) component of the velocity field, known as solenoindal velocity, and $\vec \ucomp$ is its curl-free ($\nabla \times \vec \ucomp = 0$) component, known as compressive velocity.
    Using Eq. (\ref{eq:fourier}) for any $h$ with $w_h = u_{\mathrm{tot},h}$, we obtain the Fourier coefficient of the $h$-component of $\vec \utot$, $\tilde u_{\mathrm{tot},h}$.
    We now compute the Fourier coefficient of the $h$-component of $\vec \ucomp$ as $\tilde u_{\mathrm{comp},h} = k_h \left(\sum_{i=1}^{3} k_i \tilde u_{\mathrm{tot},i}\right) / \left(\sum_{j=1}^{3} k_j k_j\right)$ and the corresponding component of $\vec \ucomp$ as $\tilde u_\mathrm{sol,h} = \tilde u_\mathrm{tot,h} - \tilde u_\mathrm{comp,h}$.
    From $\tilde u_\mathrm{sol,h}$ and $\tilde u_\mathrm{comp,h}$ for any $h$, we evaluate the 3D power spectra of $\vec \usol$ and $\vec \ucomp$ as described in Sect. \ref{sec:fourier}.
    For each subbox, we evaluate the solenoidal-to-compressive energy ratio, $\Esol / \Ecomp$, where $\Esol$ and $\Ecomp$ are the integrals of the 3D power spectra (defined in Sect. \ref{sec:fourier}) of $\usol$ and $\ucomp$, respectively.
    Given that the fluctuation dynamo is driven only by the solenoidal component of the velocity field, $\Esol / \Ecomp$ significantly contributes to determining the efficiency of the fluctuation dynamo \citep{Porter15}.
    \item The quantity $\sigma_\rho / \bar \rho$, where $\sigma_\rho$ is the standard deviation of the density distribution within the subbox, traces the presence of clumps and measures their dynamical relevance within our subbox \citep{roncarelli13, Zhuravleva13, Angelinelli21}.
    \item Radial anisotropy of the turbulent velocity dispersion tensor and skewed probability distribution functions (PDF) of the radial velocity are well known to characterize the cluster regions with the infalling substructures \citep[e.g.][]{Vazza18_support, angelinelli20}. 
    These substructures distinguish the ICM turbulence from more isotropic and volume-filling turbulence formed in fully controlled numerical experiments.
    As proxies for radial anisotropy and radial skewness of the ICM velocity field, we use the ratio of the maximum to the average, $\sigma_{\max} / \sigmamean$, and the absolute value $\gammamean$ of the average of the velocity distribution skewness along each $h$ direction within a given subbox, respectively.
    \item We quantify the effect of the bulk motions, i.e.\ the average velocity on the scale of the subbox side length, in analogy to the quantities $\mathcal M_\sigma$ and $\sigma_{\max} / \sigmamean$.
    Specifically, we evaluate $\mathcal M_u = \left( u_x^2 + u_y^2 + u_z^2\right)^{1/2} / \sound $ and $u_{\max} / u_\mathrm{mean}$, where $u_h$ is the Cartesian $h$-component of the average velocity of the ICM within a given subbox, with $h = \{x, y, z\}$.
\end{itemize}

\begin{figure*}
   \centering
   \includegraphics[width=1\textwidth]{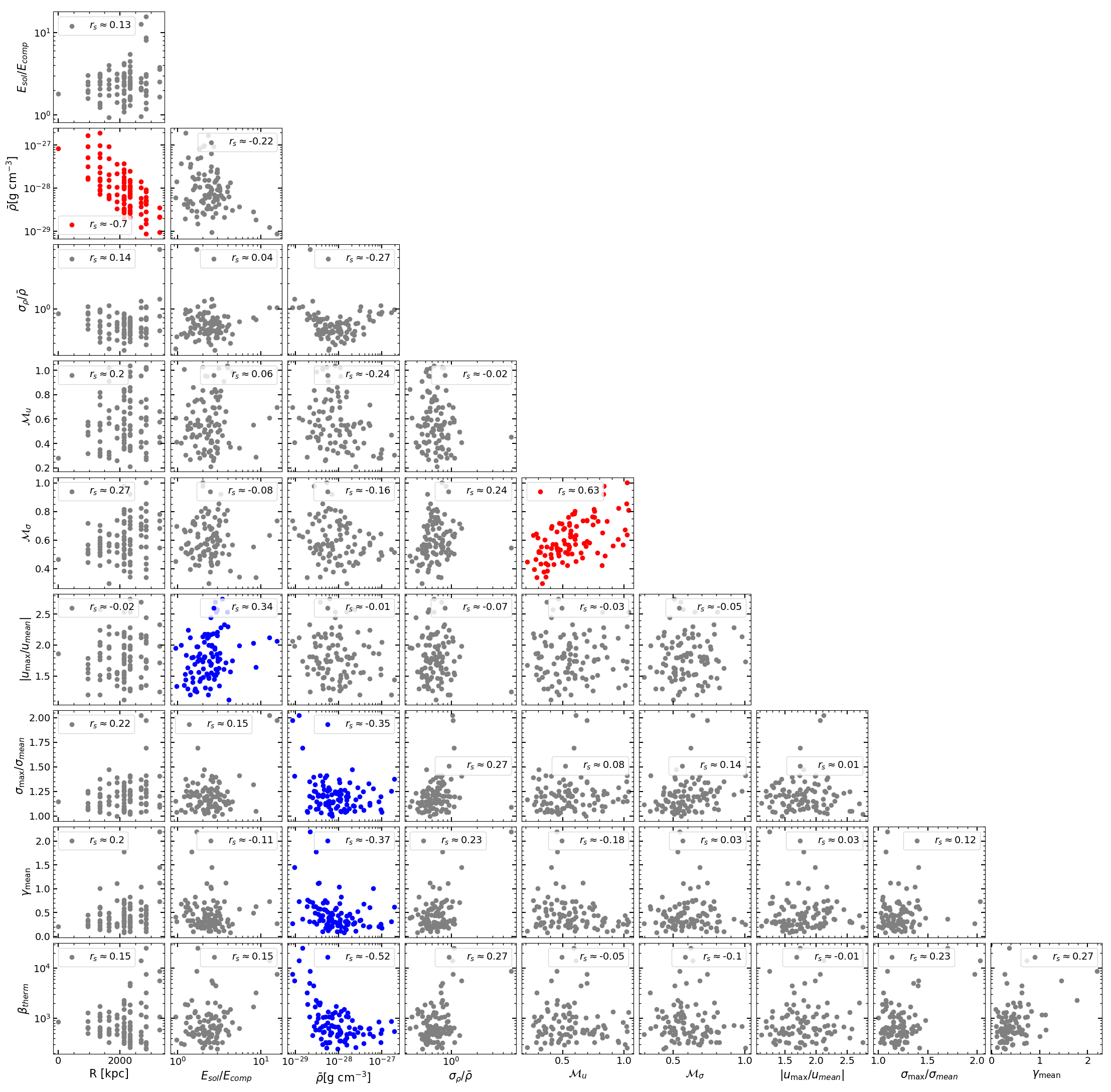}
    \caption{Scatter plots of the cross-correlations among the ICM quantities described in Appendix \ref{sec:appendix}. 
    The corresponding value of the Spearman's correlation coefficient ($\rs$) is reported in the legend of each panel.  }
    \label{fig:cross}
\end{figure*}

\begin{figure*}
   \centering
   \includegraphics[width=0.38\textwidth]{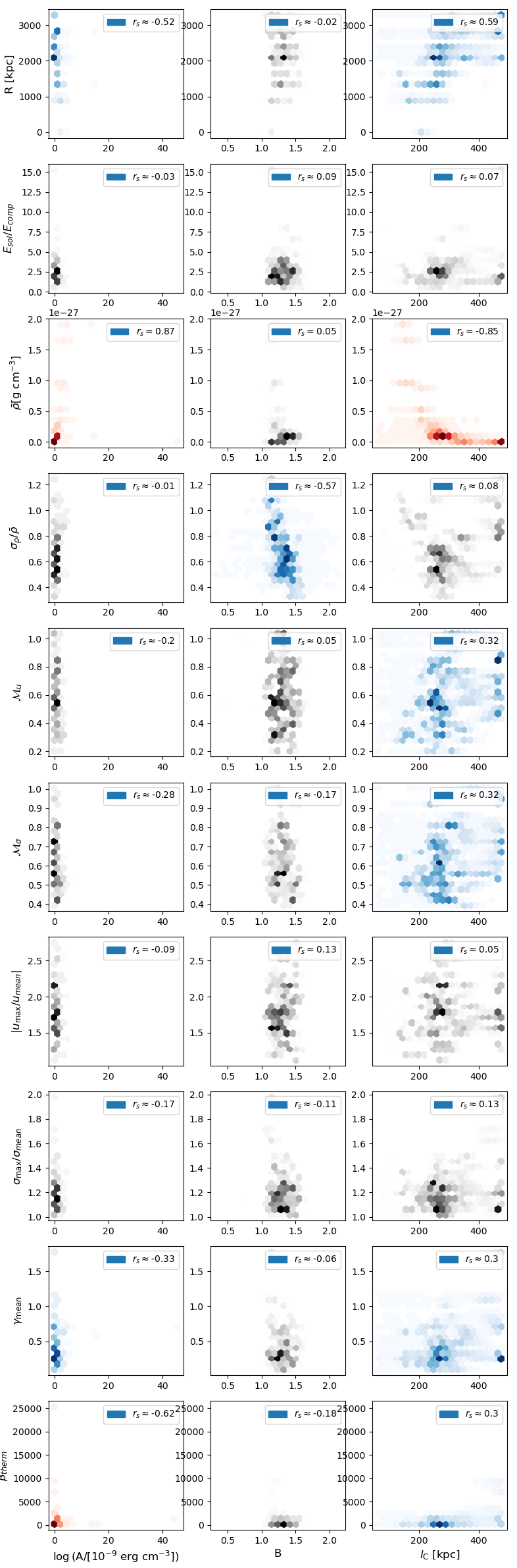}
    \caption{Scatter plots of the parameters of the dynamo-like ($D=0$) model ($A$, $B$, $\lC$; horizontal axis) and the subbox properties described in the main text (vertical axis). 
    The corresponding value of the Spearman's correlation coefficient ($\rs$) is reported in the legend. These $\rs$ are evaluated only for the $t=0.01$ subbox collection (see Sect. \ref{sec:statistics}).}
    \label{fig:corr_dynamo}
\end{figure*}

\begin{figure*}
   \centering
   \includegraphics[width=0.49\textwidth]{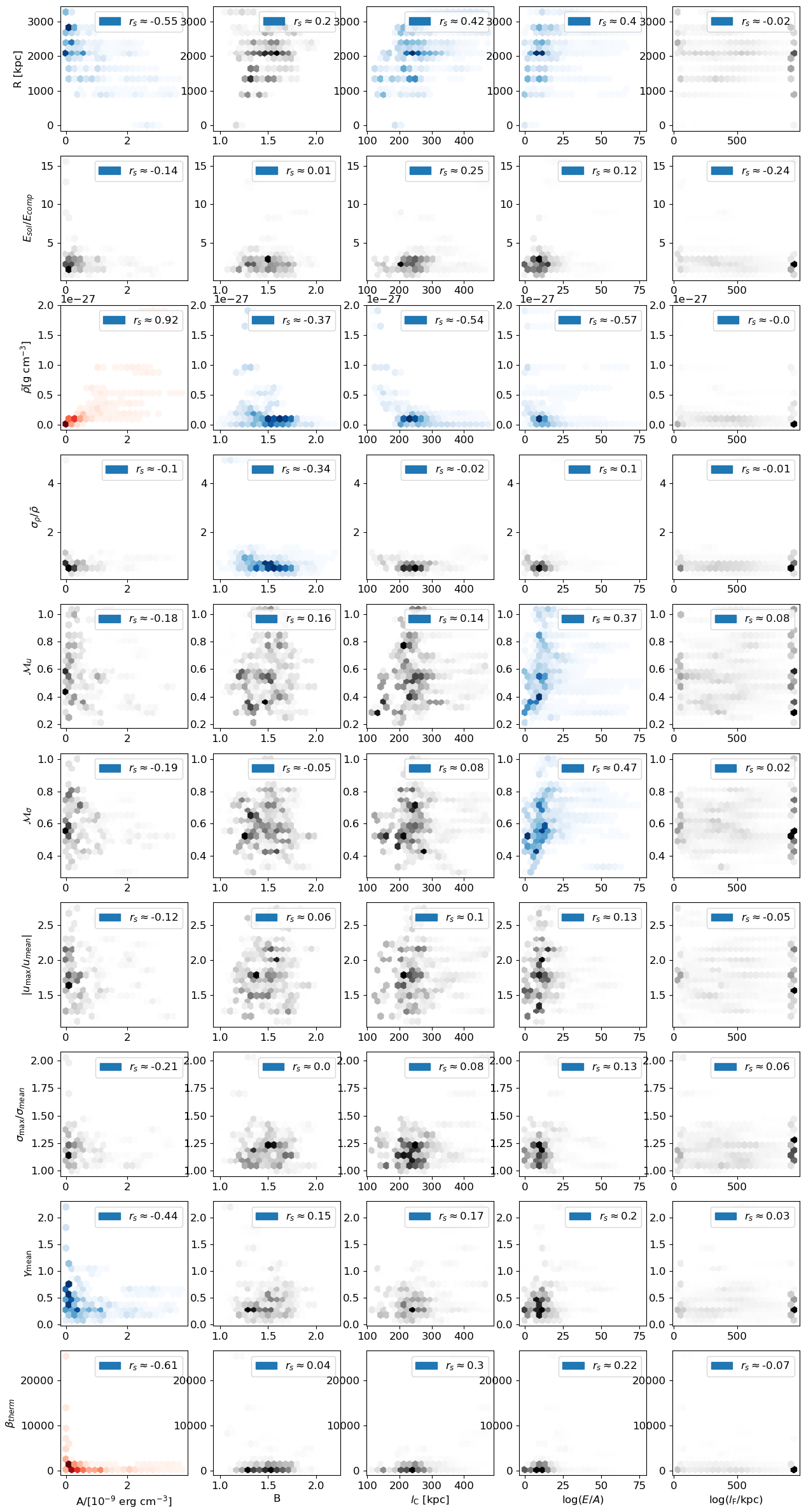}
    \caption{The same scatter plot as Fig. \ref{fig:corr_dynamo}, but for all the Eq. (\ref{eq:model}) parameters in the extended ($D\neq 0$) model ($A$, $B$, $\lC$, $D/A$ and $\lF$). }
    \label{fig:corr_extended}
\end{figure*}

Figure \ref{fig:cross} presents the cross-correlations among all the $\bar Q$ quantities introduced above in our subbox collection. 
The parameters $R$, $\sigma_{\max}/\sigmamean$, $\gammamean$, and $\betatherm$ exhibit modest or strong correlations with $\bar{\rho}$ (with $|\rs|> 0.6$), suggesting that these quantities can be approximated as being primarily stratified over the density.
If the dominant source of subbox-to-subbox variations in a given parameter $\theta$ of Eq. (\ref{eq:model}) is the density $\rho$, then $\theta$ is expected to correlate with $\rho$, with Spearman's correlation coefficient $r_{\mathrm{s},\rho}$, as well as with the aforementioned quantities $\bar Q$ that themselves depend on $\rho$. However, these latter correlations are expected to be weaker, i.e., $|r_{\mathrm{s},Q}| < |r_{\mathrm{s},\rho}|$.
In Sect. \ref{sec:correlation}, we describe a procedure to distinguish direct correlations of $\theta$, with $\rho$ in this example, from indirect correlations, arising here through the dependence of the other aforementioned quantities on $\rho$.

\subsection{Method to evaluate the degree of correlation}
\label{sec:degree}

This Section describes the method to quantify the significance of the correlation of each pair composed of one of the Eq. (\ref{eq:model}) parameters and one of the quantities reported in Sect. \ref{sec:dynamics} and presents the results of this correlation analysis.

Given that the median values of the Eq. (\ref{eq:model}) parameters, due to their estimation uncertainty and their cross-correlations, can be not in any case representative of the underlying power spectrum parameters, we associate all the 1000 $\vec \theta$ of the corresponding posterior sampling of a given subbox (see Sect. \ref{sec:statistics}) with the same $\bar Q$ and evaluate $\rs$ considering 1000 pairs for each subbox in the sample.
The uncertainty in the parameter estimation increases the noise in potential relations between model parameters and subbox properties, thereby reducing their statistical significance, %and given that a resolution of the subbox size in the evaluation of both power spectrum properties and ICM conditions can reduce $|\rs|$ of a given correlation, 
and thus we study both strong ($|\rs| \geq 0.6$) and modest ($0.3 \leq|\rs| < 0.6$) correlations.
Because two subbox properties, $\bar Q_1$ and $\bar Q_2$, may themselves be correlated, we adopt the following procedure to identify the main drivers of the parameter values.
For a fixed parameter in $\vec{\theta}$, we first identify the property $\bar Q_1$ such to yield the largest $|\rs|$ with this parameter, and denote the corresponding $\rs$ by $r_{\mathrm{s},1}$. 
We then interpret the dependence of this parameter on $\bar Q_1$ as direct.
Next, we identify the property $\bar Q_2$ associated with the second-largest $|\rs|$, denoted by $r_{\mathrm{s},2}$. 
If $\bar Q_1$ and $\bar Q_2$ are neither modestly nor strongly correlated, we also classify the dependence on $\bar Q_2$ as direct. 
Otherwise, we consider the possibility that the correlation between $\bar Q_2$ and the considered parameter can be entirely explained by the correlation between $\bar Q_1$ and $\bar Q_2$, denoted by $r_\mathrm{s,Q}$, without invoking an additional direct dependence, and we test whether the difference between $r_\mathrm{s,1}$ and $r_\mathrm{s,2}$ is statistically significant when compared to $r_\mathrm{s,Q}$. 
In the case of no significance, we classify the dependence on $\bar Q_2$ as indirect; otherwise, as direct.
The same procedure as for $\bar Q_2$ is then iteratively applied to all remaining $\bar Q$ characterized by at least a modest correlation with the considered parameter, in order to distinguish direct from indirect relations.

Figs. \ref{fig:corr_dynamo} and \ref{fig:corr_extended} show the correlation of our fitting parameters with all the aforementioned $\bar Q$ for the dynamo-like and extended models, respectively.
All the parameters except for $\lF$ ($A$, $B$, $\lC$ and $D/A$) in one of the two models at least modestly correlate with one or more $\bar Q$. 
For each of the parameters $A$, $B$, $\lC$, and $D/A$, we apply the procedure described above to disentangle direct from indirect correlations and to identify the main drivers of the subbox-to-subbox variations.
In Sect. \ref{sec:correlation} we comment on the quantities $\bar Q$ that our analysis identifies as the main drivers of the subbox-to-subbox diversity in the 3D magnetic power spectra of individual subboxes.

\end{appendix}

\end{document}